\documentclass[fleqn,usenatbib]{mnras}
\usepackage{newtxtext,newtxmath}
\usepackage[normalem]{ulem}

\usepackage[T1]{fontenc}
\usepackage{ae,aecompl}
\usepackage{float,comment}

\usepackage{graphicx}   
\usepackage{amsmath}    
\usepackage{scalerel}
\title[LD-WHIM Scaling Relation]{The Scaling Relation between 
Galaxy Luminosity and WHIM Density from EAGLE Simulations with application to SDSS data}

\author[P. Holt et al.]{
Patrick Holt$^{1}$, Toni Tuominen$^{2}$, Jukka Nevalainen$^{2}$, Massimiliano Bonamente$^{1}$\thanks{E-mail: bonamem@uah.edu},
\newauthor 
Teet Kuutma$^{2}$, Pekka Heinämäki$^{3}$, 
E. Tempel$^{2}$\\
$^{1}$Department of Physics, University of Alabama in Huntsville,Huntsville, AL \\
$^{2}$Tartu Observatory, University of Tartu, Observatooriumi 1, T{\~{o}}ravere 61602, Estonia\\
$^{3}$Tuorla Observatory, Department of Physics and Astronomy, University of Turku, Finland\\
}
\date{Accepted XXX. Received YYY; in original form ZZZ}

\pubyear{2021}

\usepackage{color,nicefrac}
\def\hst{\it HST\rm}
\def\fuse{\it FUSE\rm}

\def\ovi{O~VI}
\def\ovii{O~VII}
\def\oviii{O~VIII}

\def\hi{H~I}

\def\kmsMpc{km~s$^{-1}$~Mpc$^{-1}$}
\def\WHIM{\mathrm{WHIM}}
\def\msun{M$_{\odot}$}
\def\omegab{$\Omega_b$}
\def\OLD{$\Omega_{\mathrm{b,LD}}$}
\def\rhob{\overline{\rho_{b}}}
\begin{document}
\label{firstpage}
\pagerange{\pageref{firstpage}--\pageref{lastpage}}
\maketitle

\begin{abstract}
This paper presents an updated scaling relation between the optical luminosity density (LD) of galaxies in the $r$ band and
the density of the warm--hot intergalactic medium (WHIM) in cosmic filaments, using the high--resolution 
EAGLE simulations. 
We find a strong degree of correlation between the WHIM density and the 
galaxy luminosity density, resulting in a scaling relation between the two quantities that permits to predict the WHIM density of filaments with 
a scatter of less than $\nicefrac{1}{2}$ dex in a broad range of 
smoothed filament luminosity densities.
In order to estimate the performance of the simulation--based calibration of the LD--WHIM density relation, we applied it  
to a sample of low--redshift filaments detected with the \emph{Bisous} method in the Legacy Survey SDSS~DR12 data.
In the volume covered by the SDSS data, our relation
predicts a WHIM density amounting to 
$31\pm7\pm12$\% (statistical errors followed by systematic) of cosmic baryon density.
This agrees, albeit within the large uncertainties, with the current estimates of the cosmological missing baryon fraction, implying that our LD - WHIM density relation 
may be a useful tool in the search for the missing baryons.
This method of analysis provides a new promising avenue to study the
physical properties of the missing baryons, using an observable that is
 available for large volumes of the sky, complementary and independent from WHIM searches with absorption--line systems in the FUV or X--rays.

\end{abstract}


\section{Introduction: the need to use galaxy luminosity as a tracer of WHIM
baryons}
The diffuse intergalactic medium contains the majority of the
universe's baryons at all redshifts \citep[e.g.][]{shull2012}.
At high redshift, most of the baryons are in a photoionized phase that
is primarily detected via the Lyman~$\alpha$ forest \citep[e.g.][]{penton2000}. 
At low redshift, a large fraction of the baryons are expected to be in a warm-hot
intergalactic medium (WHIM) phase at temperatures $\log T(K) = 5-7$
\citep[][]{martizzi2018,bertone2008,dave2001,cen1999}.
The low--temperature range of WHIM temperatures
is effectively probed by the available FUV data, primarily \hst\ and \fuse,
revealing that FUV absorption lines (primarily \ovi\ and the 
\hi\ Lyman series) only trace a portion of the
low--redshift baryons \citep[][]{martizzi2018,bertone2008,dave2001,cen1999}.


\cite{shull2012} provided a popular census of the observational status of the cosmic baryon budget at low redshift,  estimating that $\approx 30 \pm 10\%$ of the baryons predicted by the concordance cosmology have not been detected. These results are later refined by \cite{danforth2016}, who estimated that the cosmic baryon density traced by \ovi\ and \hi\ with HST/COS, the most important elements detected in the FUV, is substantially lower than assumed in \cite{shull2012}. Consequently the missing baryon density may amount to $\sim$50\% of the cosmic baryon density.

There are several serious uncertainties with the estimate of the total baryons traced by the FUV (see discussion in \citealt{tuominen2021}). One key source of uncertainty is that the temperature and metal abundance of the gas are not known from observations.
The conversion into the the associated hydrogen mass is done based on simulations and it can be uncertain by a factor of several. Another problem is that \ovi\ can be detected in the same range of temperatures where \hi\  broad Lyman--$\alpha$ (BLA) lines occur, which is in the low end of the WHIM temperature range. One must therefore consider the possibility that a fraction of the baryons traced by \ovi\ are the same baryons that were detected by BLA measurements. 
Finally, it is also possible that, due to the sensitivity limits of the COS surveys, the FUV has only probed the most dense circum--galactic medium (CGM) close to the galaxies.
 FUV studies do provide evidence that at least a portion of these O~VI absorption lines 
are associated with the galaxies themselves \citep[e.g.][]{stocke2006,tchernyshyov2021}.
It is therefore possible that the warm WHIM (i.e., in the $\sim \log T(K)=5-6$ range) in the cosmic filaments has not been detected yet, and thus it constitutes an additional missing baryon component.
These considerations render the missing baryon fraction very uncertain. In this paper we adopt the commonly accepted scenario that half of the cosmic baryon budget may be missing. 

Simulations (e.g., EAGLE, \citealt{schaye2015,tuominen2021}, IllustrisTNG, \citealt{pillepich2018,gallaraga2021}) indicate that a combination of hot WHIM (i.e, $\sim \log T(K)=6-7$) and
the FUV--detectable warm WHIM may be the missing WHIM.
The high--temperature WHIM can be probed with X--ray lines such
as those from \ovii\ or \oviii, but the resolution of available
X-ray instrumentation can only provide tentative detections
of some of the more massive WHIM systems \citep[e.g.][]{nicastro2018,nicastro2018b,bonamente2016,fang2010, ahoranta2020}.
It is therefore necessary to provide alternative ways to investigate
the location and cosmological density of WHIM baryons in the local
universe.

A complementary way to investigate the missing baryons problem is by using the relatively easily obtainable optical luminosity density of galaxies in cosmic filaments as a tracer of the WHIM phase. 
The underlying physical argument responsible for the correlation between the galaxy light and WHIM density distribution is the dominating role of the gravitational potential of the dark matter concentrations at the filament spines. 
An important heating mechanism of the intergalactic baryons in the large scale structure is the adiabatic compressional heating of the matter accreted towards the filaments. The baryons falling towards the filaments also experience shock-heating which is stronger closer to the maxima of the gravitational potential, i.e. the filament spines. Thus, the intergalactic medium is expected to be the hottest 
($\log T(K) = 6-7$) within $\sim$1 Mpc of the filament spines, as demonstrated by \cite{tuominen2021} with the EAGLE data.  

The interaction of galaxies with the large scale structure is very different. Yet, the galaxy density is highest at the filament spines \citep[e.g.][]{gallarraga2020}, and
 galaxies are born at the maxima of the dark matter fluctuations. This is most likely to happen close to the filament spines, where the dark matter density is the highest. 
An initial investigation by \citet{nevalainen2015} revealed a correlation between the luminosity density of galaxies and the WHIM density in the cosmological simulations of \cite{cui2012}. The resulting scaling relation between galaxy luminosity density (hereafter LD) and the WHIM density was successfully applied to filamentary structures in the Sculptor Wall supercluster, showing that the scaling relation is consistent with the available constraints from X--ray observations of the WHIM \citep{nevalainen2015}. This preliminary study therefore provided confidence that the optical luminosity of galaxies can be used as a proxy for WHIM density.

In the present paper we update the LD-WHIM scaling relation with state-of-the-art EAGLE simulations. Our goal is to apply it to such a large  observational filament sample (from SDSS) so that an evaluation of the WHIM contribution to the cosmic baryon budget is feasible. This evaluation is necessary in order to test how well our relation performs in recovering the missing baryons in filaments. Our definition of the WHIM is gas outside r$_{200}$ of haloes, and in the temperature range of 
 $\log{T(K)}$ = 5-7.
 The paper is structured as follows.
Section \ref{sec:eagle} describes the EAGLE simulations used for the new evaluation of the LD-WHIM scaling relation. Section~\ref{sec:scalingRelation}
describes the method of analysis of the WHIM simulations to obtain
the new scaling relation. Section~\ref{sec:observations} presents our analysis of the cosmic baryon inventory by applying the LD - WHIM density relation to the SDSS filaments and Section~\ref{sec:conclusions} provides our
conclusions.

\section{EAGLE simulations}
\label{sec:eagle}
The EAGLE project (Evolution and Assembly of GaLaxies and their Environments) is a suite of hydrodynamical simulations of the evolution of the universe \citep{schaye2015}. EAGLE features the standard  $\Lambda$CDM cold dark matter and dark energy cosmology, with the simulations run in volumes of 25 to 100 comoving Mpc (cMpc) sided cubes.
The EAGLE simulations use state-of-the-art numerical techniques and subgrid models for radiative cooling, star formation, stellar mass loss and metal enrichment, energy feedback from star formation, gas accretion onto, and mergers of, supermassive black holes and AGN feedback,
 including the calibration of subgrid feedback to observables \citep{crain2015}.
The galaxy data and snapshots of the EAGLE simulations  are available to the public
\citep{mcalpine2016,eagle2017}.

EAGLE was run using a modified version of the N-body Tree-PM smoothed particle hydrodynamics (SPH) code Gadget 3 \citep{springel2005}, with the main modifications being the formulation of SPH, the time steps, and the subgrid physics. Of the available EAGLE simulations, we use the largest simulation (Ref-L0100N1504)
at a redshift of $z=0$.
The simulation has a comoving box size of $100$ cMpc, a total of $2\times 1504^{3}$
particles, an initial baryonic particle mass of $1.81\times 10^{6} M_\odot$, a dark matter particle mass of $9.7 \times 10^{6} M_\odot$, comoving Plummer-equivalent gravitational softening length of $2.66$~ckpc, and the maximum physical softening length value of $0.70$~kpc.
This is the same EAGLE simulation used by \cite{tuominen2021} for the investigation of the missing baryons in the local universe.

\subsection{Constructing the WHIM Density and Temperature Grid}
\label{grid}
The baryon particles in the simulation, along with the associated mass, temperature, and density for each particle, were used to create a $500^3$ uniform grid that forms the basic dataset we use for this study. Since our interest is in the diffuse warm--hot intergalactic medium,
we excluded particles in bound structures, namely those within $R_{200}$ of
a virialized structure as identified 
by a friends--of--friends halo given in the EAGLE database \citep{mcalpine2016}.
Each grid point is therefore representative of a cubic cell of volume $V=0.2^3$ Mpc$^3$
with a mass equal to the sum of the masses of all particles outside 
$r_{200}$ within the 
$j$--th cell,
\begin{equation}
M_j=\sum_{i} m_{i}(r>r_{200}).
\end{equation}
Using the temperature of each particle not within a bound structure, 
we calculated the mean mass--weighted weighted temperature for each cell $j$,
\begin{equation}
T_j=\dfrac{1}{M_{j}} \sum_{i} m_{i}(r>r_{200}) \cdot T_{i}(r>r_{200}).
\end{equation}
To evaluate the density of the intergalactic WHIM, we need to estimate the portion of the
volume of the cell occupied by virialized structures.
For this purpose, we take the mass $m_{b,i}=m_{i}(r\leq r_{200})$ of each bound particle and divided it by the particle density $\rho_{b,i}$, so that  the volume occupied
by bound structures in each cell is
\begin{equation}
V_{b,j}=\sum_{i}\frac{m_{b,i}}{\rho_{b,i}}.
\end{equation}
Next, we calculate the WHIM density for each grid point using all particles, and selecting cells with an average temperatures within the WHIM range, $\log T(K)=5-7$, resulting in
\begin{equation}
\rho_{j,\WHIM}=\frac{M_{j}}{V-V_{b,j}}.
\end{equation}
 To complete the construction of the WHIM density cube, we have to exclude temperatures outside the WHIM temperature range, which we chose
 as $\log T(K)=5-7$. For gridpoints with $T_j$ values outside the WHIM temperature range, we set the mass density equal to zero.
 Figures~6 and 5 in \cite{tuominen2021} show examples of the WHIM density and temperature map in the EAGLE simulations obtained following this method. One of the advantages of mapping the EAGLE simulations into a regular grid is the ease of comparison with observational data, as shown in Sect.~\ref{sec:observations}.

\subsection{Galaxy filaments}
We adopted the EAGLE filaments of galaxies identified in \cite{tuominen2021} and \cite{kuutma2020} 
using the \textit{Bisous} method devised by \cite{stoica2007} and successfully applied to various numerical simulations \citep{tempel2014c,muru2021,nevalainen2015}. The Bisous method 
uses connected cylindrical volumes to identify coherent galaxy structures that can be identified as filament network. The current implementation of the Bisous method models the galaxy distributions with connected cylinders of a relatively narrow range of radii (0.5-1.0 Mpc), the EAGLE sample average being at $\sim 0.75$~Mpc. 
The process was repeated 1,000 times to account for the stochastic nature of the Bisous formalism. 
The locations that are more often covered by the cylinders in the individual runs are more likely to belong to a filament.
At the locations of the filament spines, the rate of coverage by the cylinder reaches the local maxima. 
The boundaries of a given filament, which we need in order to capture the diffuse LD and WHIM within filaments, are drawn at a location around the filament spine where the rate of the cylinder coverage decreases below an experimentally set limit, chosen in such a way that
it yielded the statistical properties of the SDSS filaments (see \citealt{tempel2014a} for additional details).

\cite{tuominen2021} showed that these choices result in EAGLE
filaments that have a typical radius on 1~Mpc.
The main statistics
of the filaments, LD and WHIM for the entire EAGLE volume are provided in Table~\ref{tab:EAGLE}.

\subsection{Constructing the Luminosity Density Grid}
\label{sec:LD}
The primary goal of this study is to use the galaxy luminosity
as a proxy for the WHIM density field. 
For this purpose we follow closely the procedure described in \cite{tuominen2021} to generate a smoothed luminosity density (LD)
field with the positions and $r$--band luminosities of all EAGLE galaxies with $M_r \leq -18.4$.  We adopted the galaxy sample and luminosities from \cite{kuutma2020}, i.e. for each galaxy we used the magnitude with dust attenuation as described in \citet{2018ApJS..234...20C}
if such value was available. Otherwise we used the dust-free magnitudes.
 This approach yielded an average  $r$--band LD of $<\mathrm{LD}>=7.1 \times 10^7$~L$_{\odot}$~Mpc$^{-3}$ \citep{tuominen2021}.

Such a choice of $M_r$ in EAGLE mimics the number density of SDSS galaxies with a limiting magnitude of $M_r = -19$ at $z=0.05$. In order to recreate a comparable filamentary network, it is essential to use galaxy samples with the same number density of galaxies. Using the same magnitude cut for EAGLE and SDSS would leave the former with a lower density of galaxies, changing the properties of the obtained filaments. Since $M_r = -19$ for SDSS galaxies effectively limits the distance up to where a complete filament network has been constructed for $\textit{Bisous}$ \citep{tempel2014a}, we selected $M_r \leq -18.4$ for EAGLE to produce similar filaments.

\begin{figure*}
    \centering
    \includegraphics[width=6in]{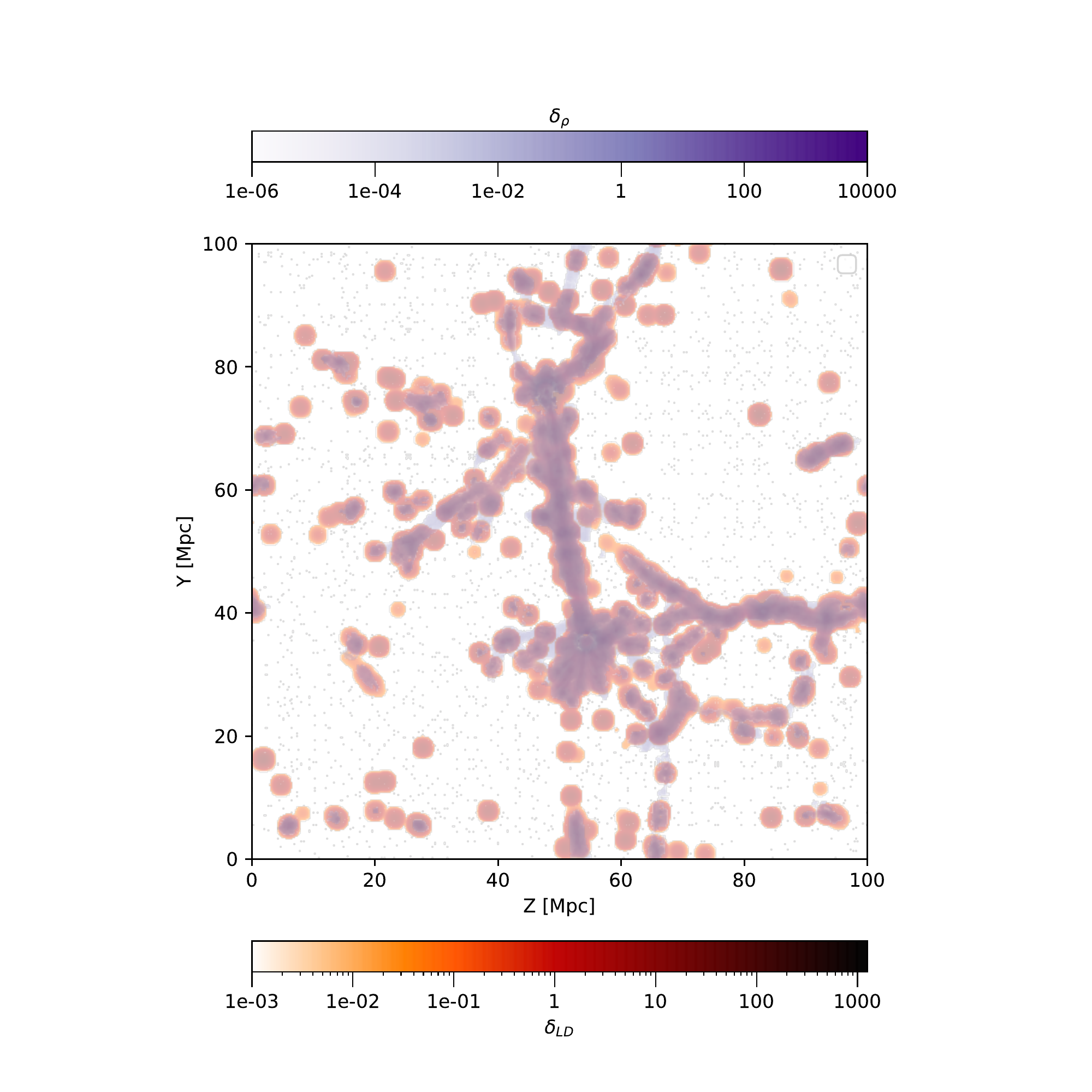}
       \caption{Image of 5 Mpc thick slice of the WHIM and LD within temperature range $T=\log 5-7$. Blue colors trace the WHIM, and 
       red colors trace the smoothed luminosity density of galaxies. } 
    \label{fig:LD-WHIM}
\end{figure*}

 A necessary step to compare the diffuse WHIM with the galaxy luminosity
 is to smooth the LD distribution on scales that are comparable to those
 of the WHIM. For this purpose, we continue to follow
 \cite{tuominen2021} and 
 smooth the galaxy luminosity with 
  a $B_3$ spline kernel
 \citep{liivamagi2012}. We then evaluate the continuous LD field in cells with a regular grid of size 
 0.2~Mpc on each side, same as for the WHIM density of Sect.~\ref{grid}. An example of a LD grid overlaid on the WHIM density fields is shown in 
 Figure~\ref{fig:LD-WHIM}, highlighting the spatial overlap between the smoothed LD and the WHIM density that it is intended to trace.

\begin{table*}
    \centering
    \begin{tabular}{l|l}
    \hline
         Number of cells &  125,000,000\\
         Total Volume   & $100^3$ Mpc$^3$\\
         Individual Cell Volume    & $0.2^3$  Mpc$^3$ \\
         Number and fraction of cells in \textit{Bisous} filaments & 6,267,543 (5.0\%)\\
         Number and fraction of cells in WHIM ($5 \leq \log T(K) \leq 7$)  & 4,263,052 (3.3\%)\\
         Average LD in EAGLE &  $7.1\times 10^{7}$ $L_{\odot}$ Mpc$^{-3}$ \\
         WHIM mass in filaments & $2.15 \times 10^{15} M_\odot$  \\
         Baryon fraction of WHIM in {\bf \emph{Bisous}} filaments & $\Omega_{\mathrm{WHIM,fil}} = 0.35$ \\
         Hot WHIM ($6 \leq \log T(K) \leq 7$) mass in {\bf \emph{Bisous}} filaments & $0.86 \times 10^{15} M_\odot$ \\
         \hline 
         Adopted baryon density & $\Omega_b = 0.05$ \\
         Adopted Hubble constant (from \textit{Planck}) & $H_0 =67.8$ \kmsMpc \\
         Mean baryon density (from \textit{Planck}) &
                $\rhob =  
                \dfrac{3 H_0^2}{8 \pi G} \Omega_b=0.618\times 10^{10} 
                \textrm{M}_\odot \textrm{Mpc}^{-3}$ \\
    \hline
    \end{tabular}
    \caption{Main statistics of filaments, WHIM and LD properties in the 
    EAGLE simulations at $z=0$, and associated cosmological quantities. Throughout this paper we use the \textit{Planck} cosmological parameters that are presented in \protect\cite{planck2013-cosmology}.}
    \label{tab:EAGLE}
\end{table*}


\begin{figure}
    \centering
    \includegraphics[width=3in]{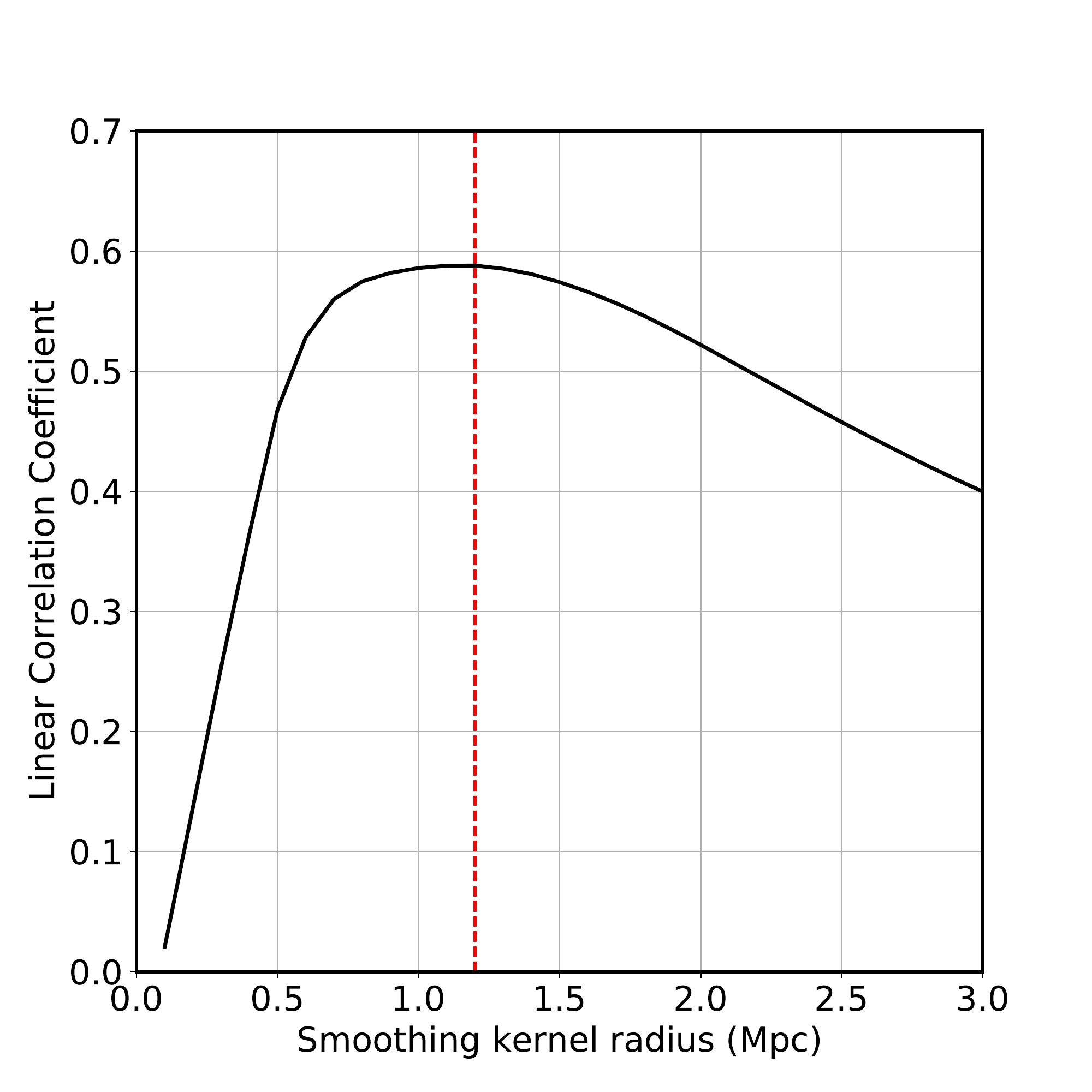}
    \caption{Linear correlation coefficient between the logarithms of the luminosity density and WHIM density, as a function of LD smoothing kernel radius $a$. The maximum is for $a=1.2$~Mpc, marked by the dashed red line.  }
    \label{fig:rvsa}
\end{figure}

\begin{figure}
    \centering
    \includegraphics[width=3.5in]{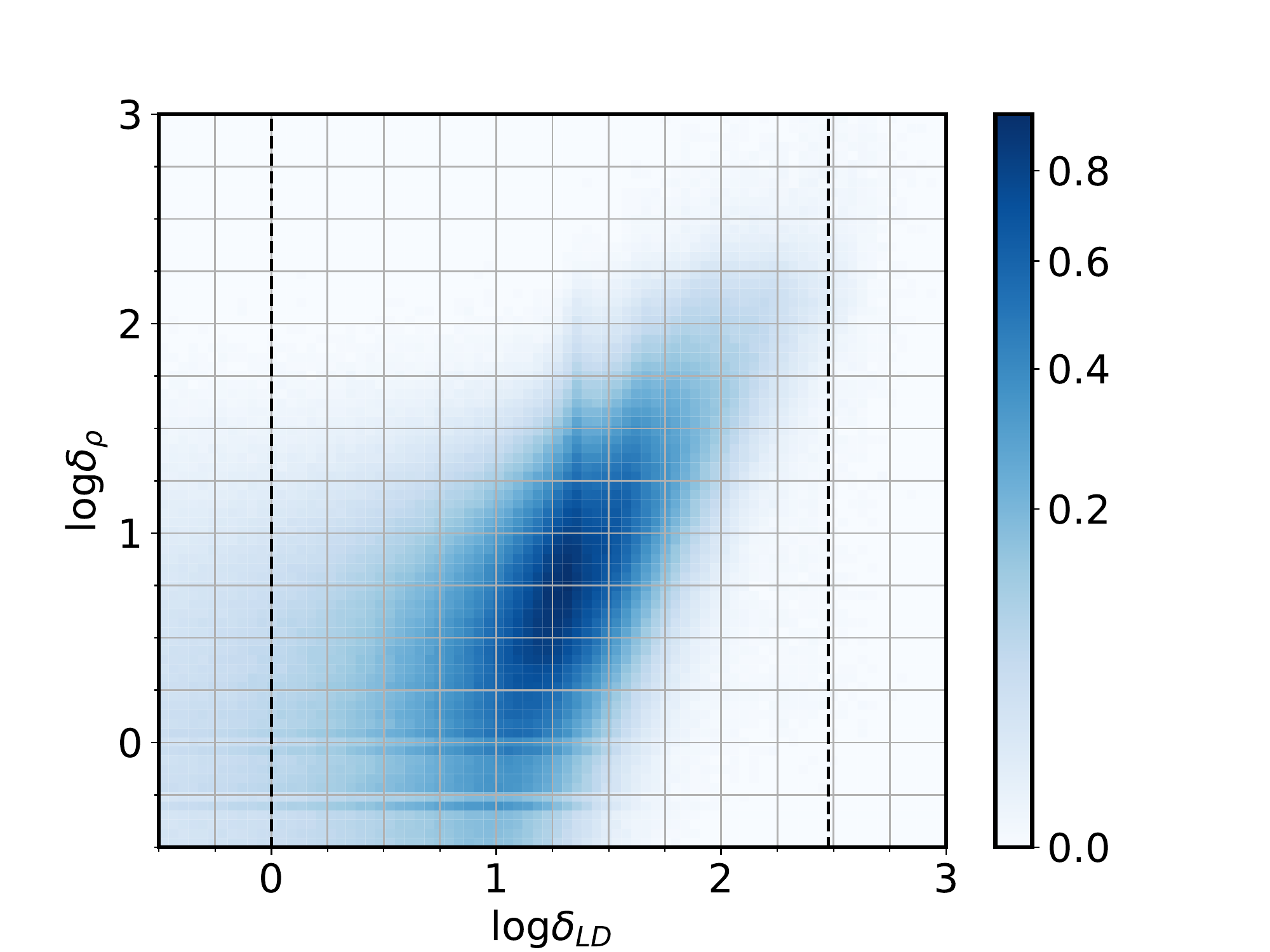}
    \caption{Distribution of LD overdensity and WHIM overdensity for $a=1.2$~Mpc, illustrating the positive correlation between the two quantities. Darker colors indicate larger frequency of occurrence (numbers are frequency density). The vertical lines indicate the LD range we selected for the further analysis $\delta_{LD}=1-300$.
    }
    \label{fig:scatter}
\end{figure}

\section{The LD-WHIM scaling relation}
\label{sec:scalingRelation}
\subsection{Defining the scaling relation}
Following  \cite{nevalainen2015}, we seek a relation between the LD and the WHIM density. 
We do this in terms of the luminosity overdensity,
defined as 
\[ \delta_{LD} = \frac{\mathrm{LD}}{\langle \mathrm{LD} \rangle} \]
where $\langle \mathrm{LD} \rangle$ is
the mean luminosity density in the r-band of the full EAGLE simulation,
and the 
and WHIM overdensity,
\[
\delta_{\rho} = \frac{\rho_{\mathrm{WHIM}}}{\rhob}
\]
where $\rhob$ 
is the mean cosmic baryon density (see Table~\ref{tab:EAGLE}).
In \cite{nevalainen2015}, the model chosen for the LD-WHIM relation was a power--law model
\begin{align*}
    \delta_{\rho} = A\, \delta_{LD}^{B}
    \label{eq:powerlaw}
\end{align*}
featuring $A=0.7 \pm 0.1 $ and $B=0.9\pm0.2$ as the best--fit values and standard deviations of the two adjustable parameter. 

In this paper we choose to work with the logarithm of the LD and WHIM density data. On a logarithmic scale, the same power--law model becomes a linear model.
\begin{equation}
\log \delta_{\rho} = \log A + B\, \log \delta_{LD}
\label{eq:logSR}
\end{equation}
where the free parameter are now $\log A$ and $B$, the latter remaining the same between our 
analysis and that of \cite{nevalainen2015}.
The use of the linear model \eqref{eq:logSR} requires the
transformation of the LD and the WHIM density to logarithmic scales before the fit, and therefore
the exclusion of cells with zero values.
As discussed below, the LD of interest is $\delta_{LD}>1$, and therefore only a negligible
fraction of cells with $\delta_{\rho}=0$ are excluded.

\subsection{Finding the optimal LD smoothing scale}
\label{sec:rvsa}
%
%
Our Bisous method is tuned to detect the galaxy filament network at $\sim$ 1--2 Mpc width scales, since it is expected that this is the characteristic scale of the dominant galaxy filaments
{\bf (see, e.g., \citealt{tempel2014a,tempel2014b,tempel2014c,tempel2016}.)}
However, the gas and galaxies are different physical entities and thus it is not obvious that the value of the optimal LD smoothing kernel radius equals the average Bisous cylinder radius of 0.75 Mpc. 
Thus, to optimise our methods, our task is to find what LD smoothing scale yields the best 
correlation between the smoothed LD and the WHIM distribution in the filaments detected with our current Bisous setting.

For this purpose we start by generating 50 smoothed 
LD fields using a range $a=0.1-5$~Mpc of smoothing kernel radii, with a step size of $0.1$~Mpc. 
We then computed the Pearson linear correlation coefficient between the LD grids and the WHIM density grid, as a function of $a$.
Since our interest is in establishing a power--law relationship between LD and
WHIM density, as defined in \eqref{eq:logSR}, the appropriate measure of the degree of
correlation is the linear coefficient between the
logarithm of the two quantities.
In fact, the Pearson coefficient is equal to 
the product $r^2=b \cdot b'$, where $b$ and $b'$ are the slopes of the linear regression of y on x, and of x on y, respectively \citep{bonamente2017book}. It is therefore appropriate to evaluate
the correlation coefficient for the logarithm of the quantities,
since the scaling relation of interest is a linear function
in the logarithms of the two quantities, see Eq.~\ref{eq:logSR} above. The results of this analysis are shown in Fig.~\ref{fig:rvsa}.

At the smallest values of the smoothing kernel radius, the light is scattered to very small distances from the galaxies. Since the diffuse WHIM fills the filaments well, most of the WHIM in filaments will not be traced with LD (i.e., many of the cells with large WHIM density have zero LD). This is reflected to a low correlation coefficient (see Fig. \ref{fig:rvsa}). The increasing smoothing scale increases the correlation coefficient, which peaks at $a \sim 1.2$~Mpc (see Fig. \ref{fig:rvsa}), i.e. at a somewhat higher scale than the average cylinder radius of 0.75 Mpc used for modelling the galaxy distribution when detecting the filaments with the Bisous method. The correlation is highly significant.
At higher scales, an increasing fraction of the light is scattered towards WHIM-poor voids, leading to a lower correlation coefficient. 
We thus adopted $a=1.2$~Mpc as the LD smoothing kernel radius.
We will evaluate the resulting scatter of the traced WHIM in Section \ref{sec:LDScalingRelation} obtained with this choice. Application of a different kernel radius would yield a lower correlation coefficient, i.e., lead to a larger scatter between the LD and WHIM.  Our analysis therefore
suggests the use of the $a=1.2$~Mpc kernel, and that 
significantly different values of the kernel size will reduce
the correlation between the WHIM density and the luminosity
density.



The two--dimensional distribution of the logarithms of the WHIM overdensity and
luminosity overdensity are shown in Fig.~\ref{fig:scatter}, for the adopted
smoothing parameter $a=1.2$~Mpc.
As expected, the higher LD corresponds to higher WHIM densities, and the 
strong degree of positive correlation between the two quantities is
evident. The correlation is less evident at the lowest densities.
Typically, cells with baryon densities that are just 
a few times the cosmic average are found close to the filament boundaries, 
approximately at a 1~Mpc distance from the filament spines (\citealt{tuominen2021}, Fig.  15).  Beyond the filament boundary the baryon density drops to underdense levels, and at a distance of $\sim$4 Mpc the WHIM densities reach on average the voids which are underdense by a factor of $\sim$10. Since the filament boundaries don't have a fixed radius,
smoothing with a fixed--radius kernel with $a=1.2$~Mpc results in
the possibility that a fraction of the luminosity density falls beyond the boundary of a thinner filament, thus leading to a larger scatter in the correlation
between WHIM density and the smoothed galaxy luminosity density, and therefore
the lower correlation coefficient observed in Fig.~\ref{fig:scatter} at the smallest values of the luminosity overdensity.
On the other hand, cells with a density contrast $\delta_{LD}=10-100$ are closer to the filament spines, and there the data  do not suffer from the boundary effect;
as a result, the correlation is stronger (see top--right portion of the distribution in Fig.~\ref{fig:scatter}).




\subsection{The LD-WHIM scaling relation from EAGLE}
\label{sec:LDbins}
\label{sec:LDScalingRelation}

\begin{figure}
\centering
        \includegraphics[width=3.4in,angle=0]{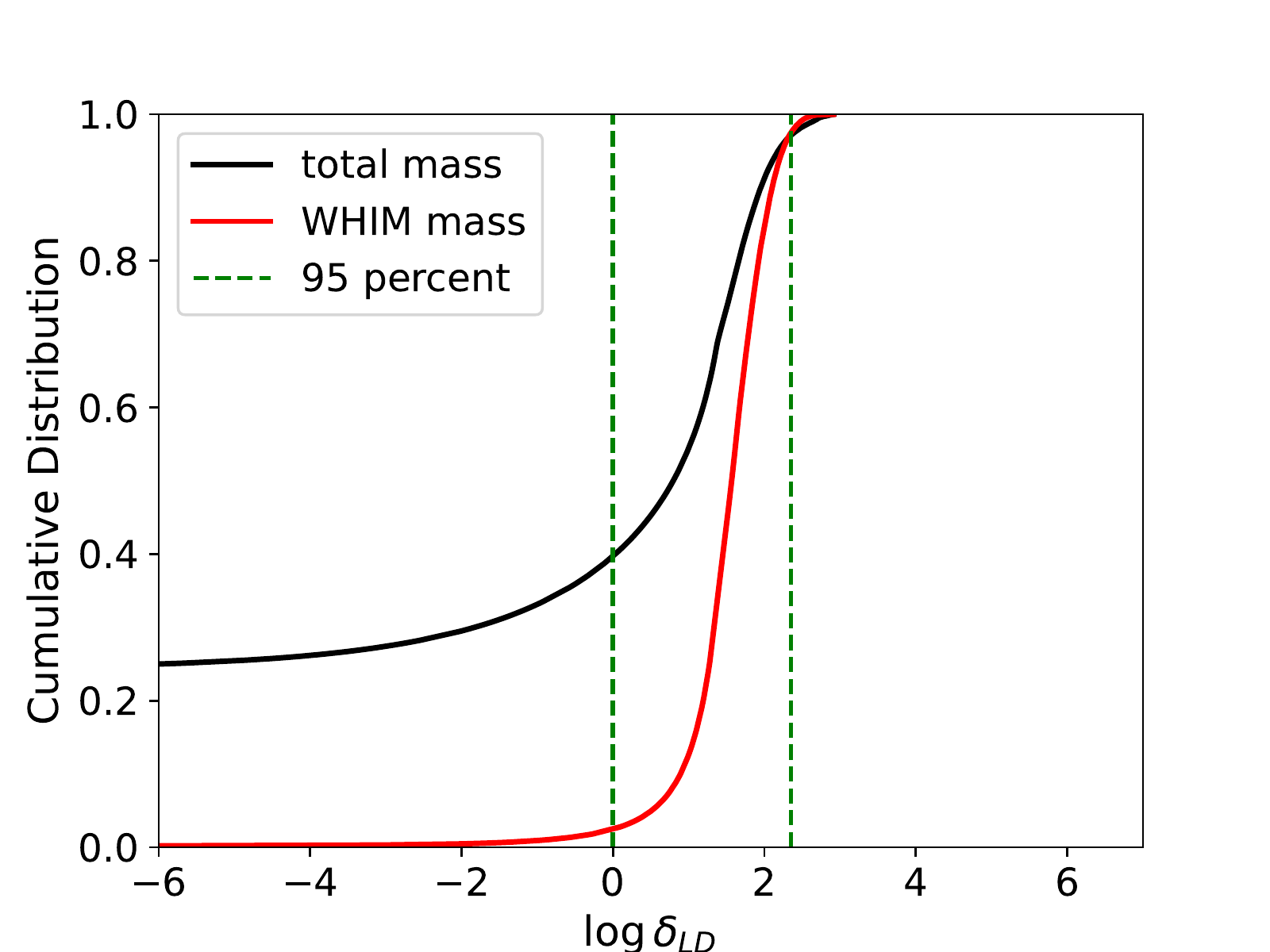} 
        \caption{Cumulative distribution of WHIM mass as a 
        function of the luminosity overdensity, for $a=1.2$ Mpc,  with
        the 95\% range indicated by green solid lines. For comparison, we also plot the distribution of mass of all the intergalactic medium, whith $\sim$25\% of its mass in cells of $\delta_{LD}=0$. }
        \label{fig:CumulativeLD}
\end{figure}

To determine the optimal LD interval range to be used in the the scaling relation,
we evaluated the  cumulative distribution of the WHIM mass in filaments
as a function of the luminosity overdensity $\delta_{LD}$, shown in Figure~\ref{fig:CumulativeLD}
for $a=1.2$. For this value of
the smoothing parameter, 95\% of the WHIM mass is included in cells
with a luminosity overdensity of approximately $\delta_{LD}\simeq 1-224$.
The luminosity overdensity range for other smoothing values are shown in Table~\ref{tab:LDpercent}.
The range of luminosities chosen by \cite{nevalainen2015} to encompass 95\% of the
mass in their simulations, i.e., 0.01-2~$\times 10^{10}$~L$_{\odot}$~Mpc$^{-3}$, 
corresponds to a very similar range of luminosity overdensities, indicating a broad
agreement with the earlier results.
Since (a) we are not interested in underdense voids and (b) 
Figure~\ref{fig:CumulativeLD} shows that there is very little WHIM at the highest densities (i.e., outside the virial regions of galaxies), we choose a luminosity overdensity range of $\delta_{LD}=1-300$ for the evaluation of the scaling relation.
The uncertainties associated with this choice are also evaluated in Sec.~\ref{sec:systematics}.
 
\begin{table*}
    \centering
    \caption{Range of the logarithm of the luminosity overdensity $\delta_{LD}$ that include 95\% of the WHIM mass. For comparison it is also reported the percent of mass included in the LD limits of \protect\cite{nevalainen2015}.}
    \begin{tabular}{c|ccccc}
    \hline
        $a$  & \multicolumn{2}{c}{95 Percentiles ($\log \delta_{LD}$)} & \multicolumn{2}{c}{No. zeros in LD} & \%\ mass in LD=0.01-2 \\
           & \multicolumn{2}{c}{\hrulefill} & Filaments & Whole Box \\
        \hline
        1.0 & -0.85 &  2.44 & 390361 & 99727119 & 0.93\\
        1.2 & -0.006 & 2.35 & 147129 &  88419657 & 0.96 \\
        1.5 & 0.4 &  2.24  & 25627 & 70396352   & 0.98\\
        2.0 & 0.55 & 2.09 & 271 & 42720903     & 0.995\\
        \hline
    \end{tabular}
    \label{tab:LDpercent}
\end{table*}

\begin{figure*}
    \centering
    \includegraphics[width=3in]{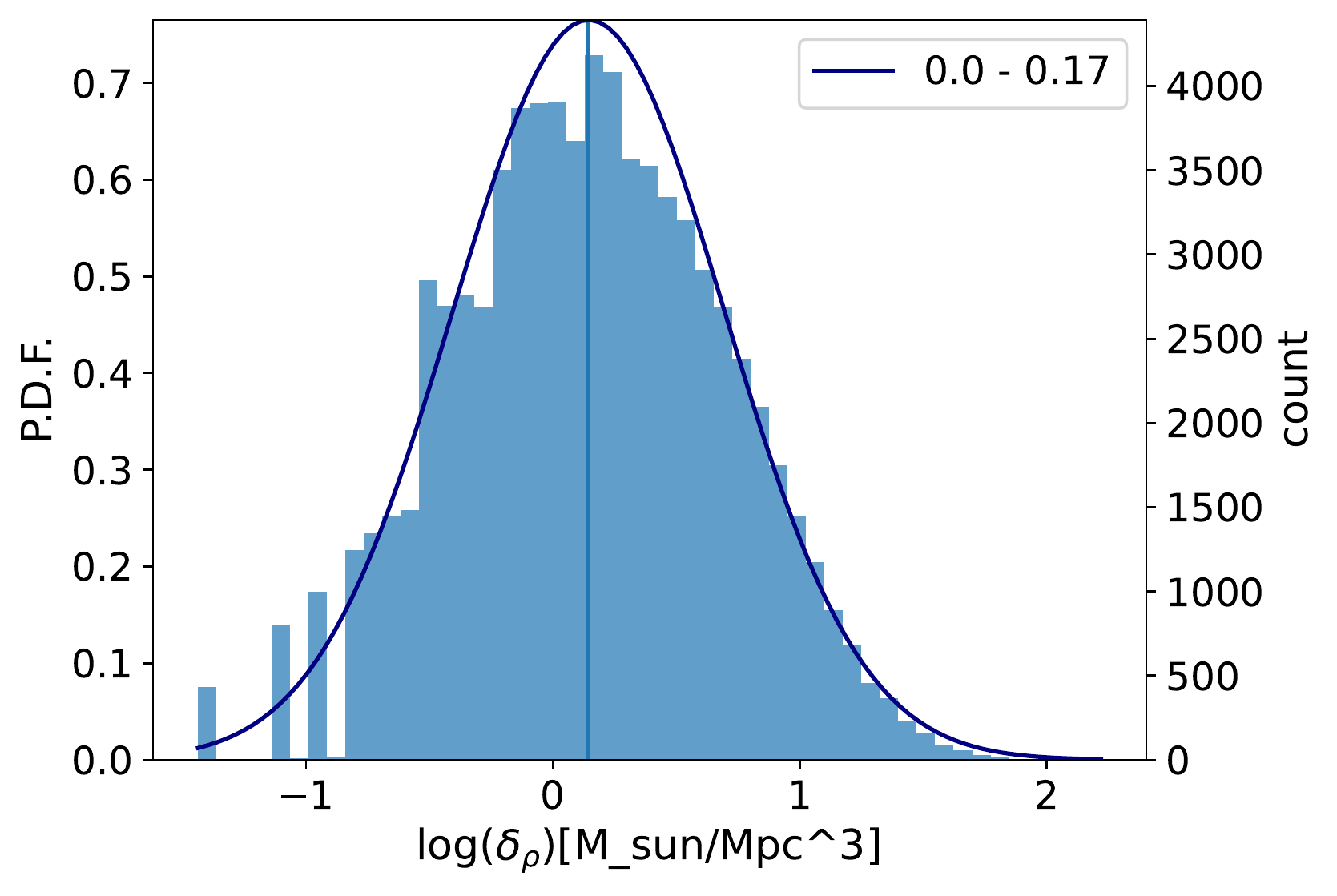}
    \includegraphics[width=3in]{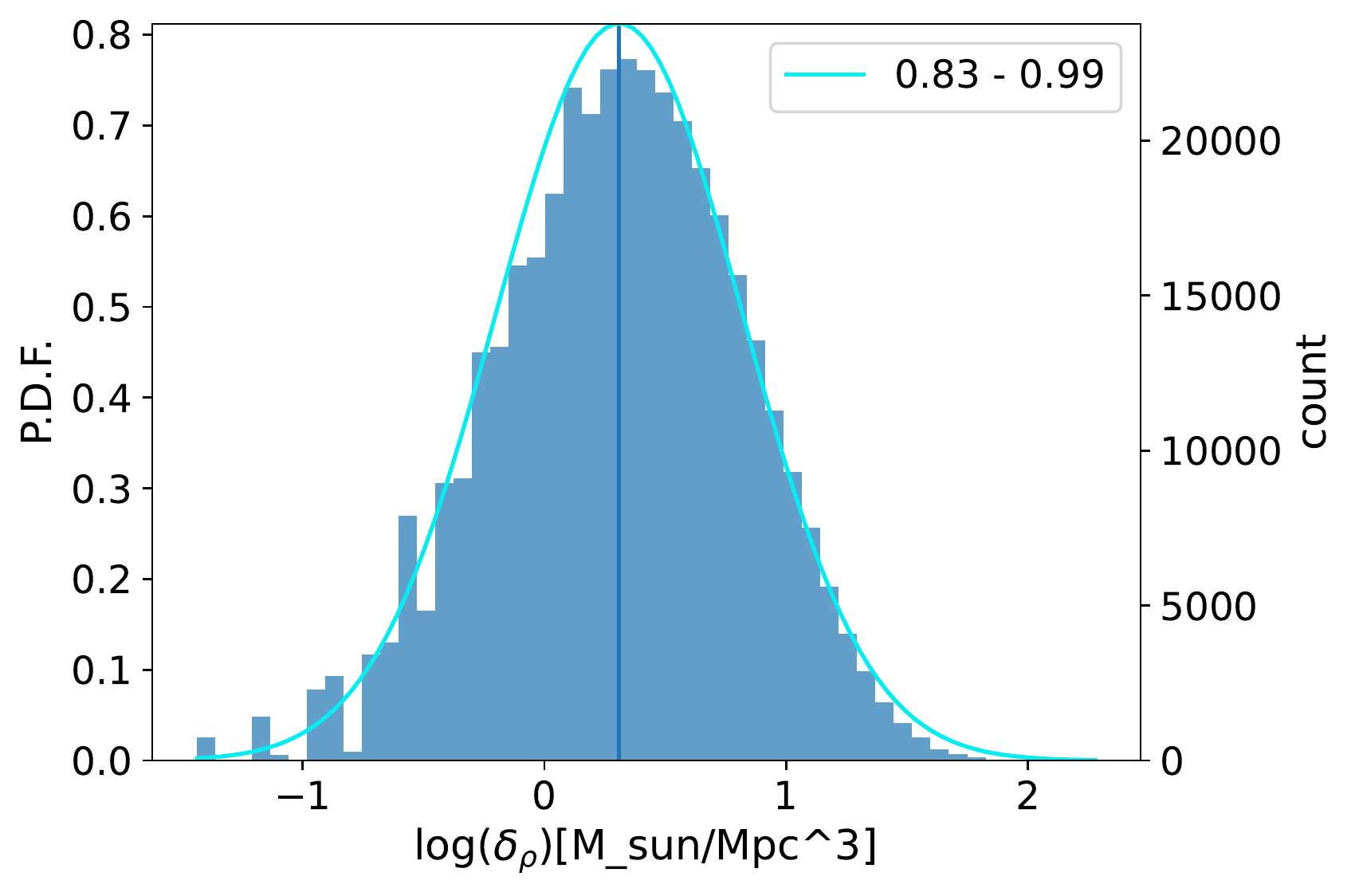}
    
    \includegraphics[width=3in]{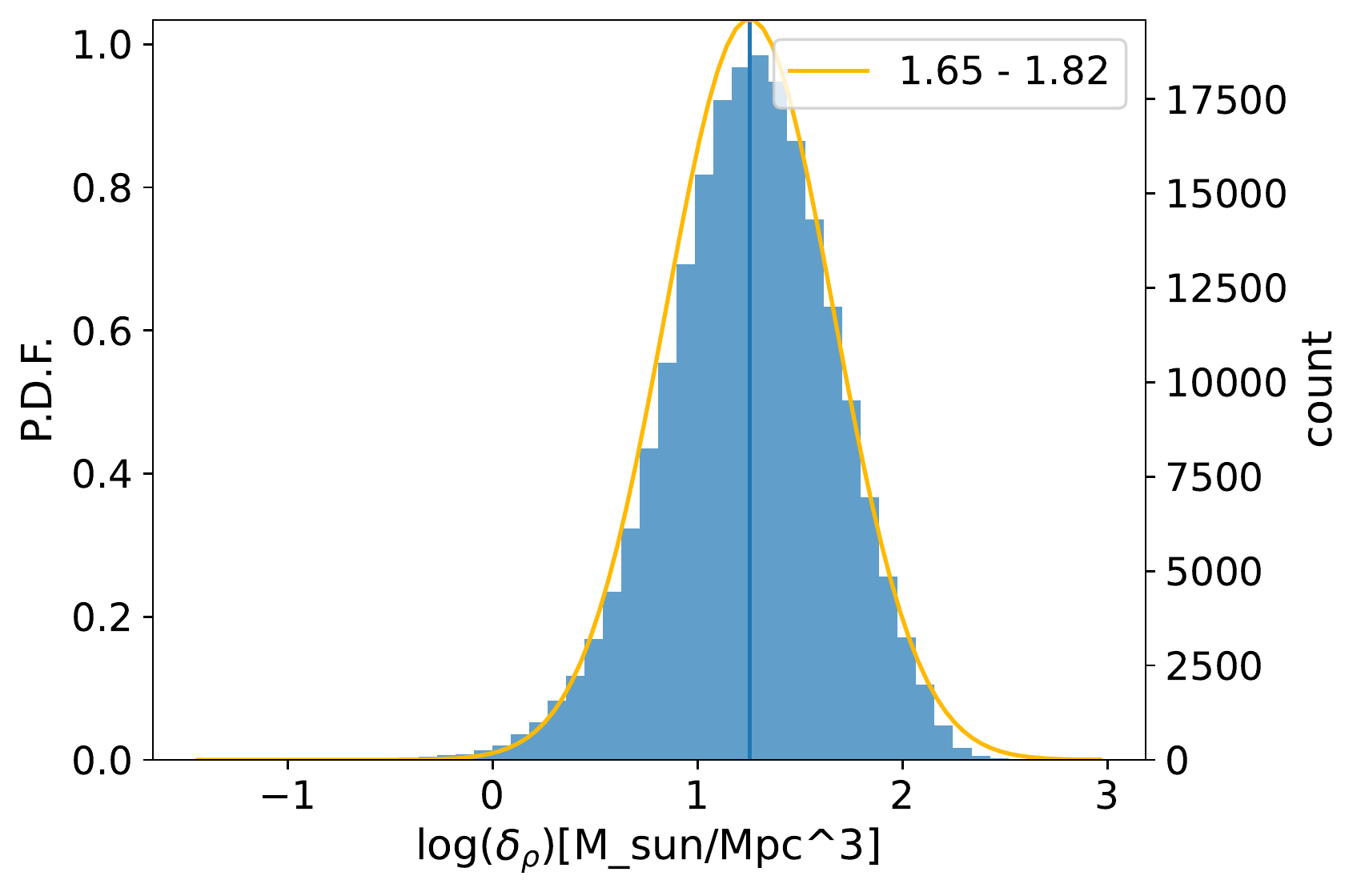}
    \includegraphics[width=3in]{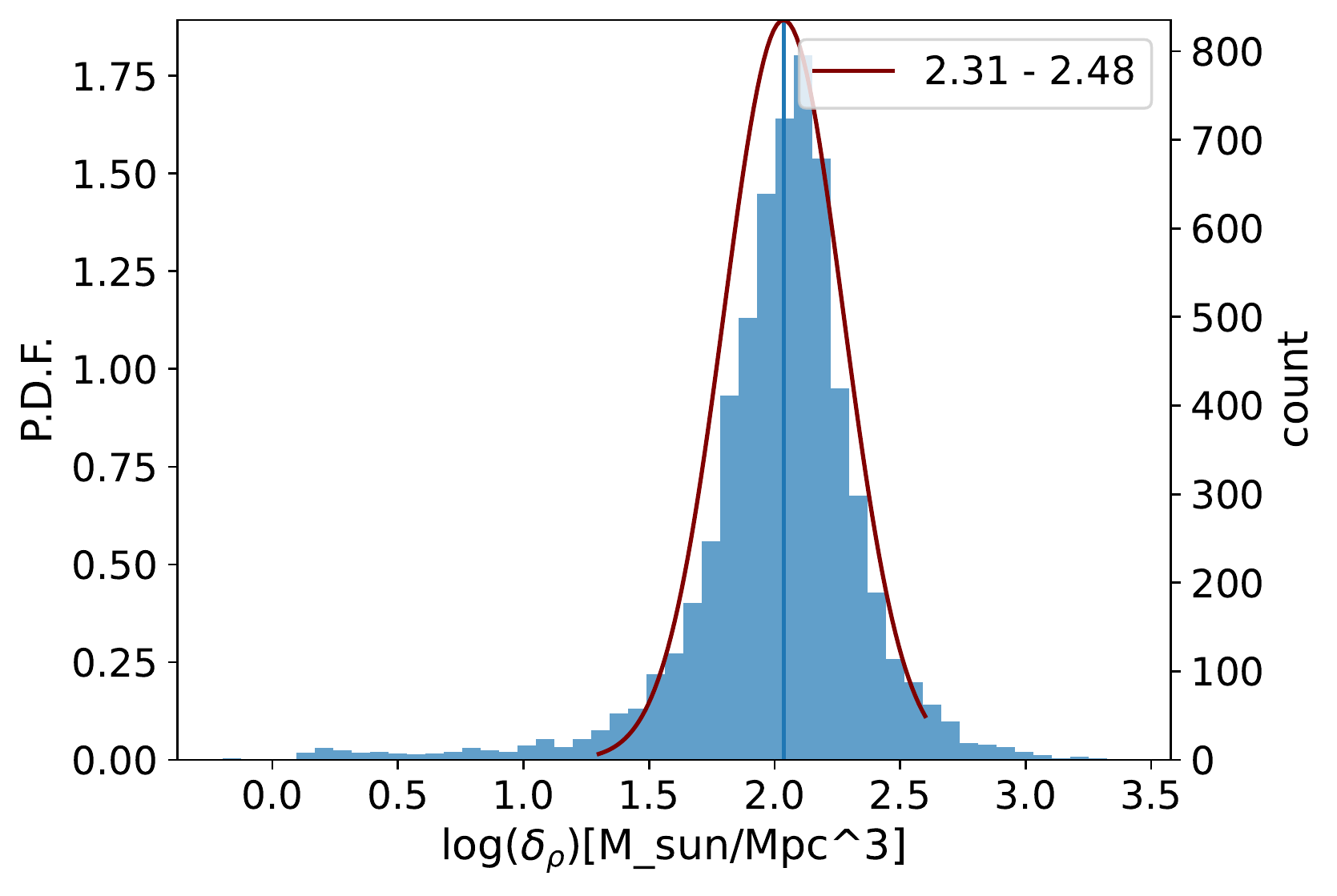}
    \caption{Sample distribution of WHIM overdensity accumulated within the $\log \delta_{LD}$ range indicated in the top right corner of the panel. 
      For the two bins with the largest LD values, the distribution has tails that are not well described by the lognormal model, as shown in the bottom--right panel. For these two distributions we chose
    to exclude the extreme portions of the tails of the distributions in order to provide a reasonable approximation to the bulk
    of the WHIM density values with the lognormal model (the range that was excised was chosen ad--hoc, where the bins contained only few counts; the range in use corresponds to the support of the solid curve). For the lowest--LD bins, the left wing of the distribution has sparse coverage, in correspondence to the horizontal stripes visible also in Fig.~\ref{fig:scatter}.}  
    \label{fig:hist1}
\end{figure*}

\begin{figure*}
\centering
        \includegraphics[width=6in,angle=0]{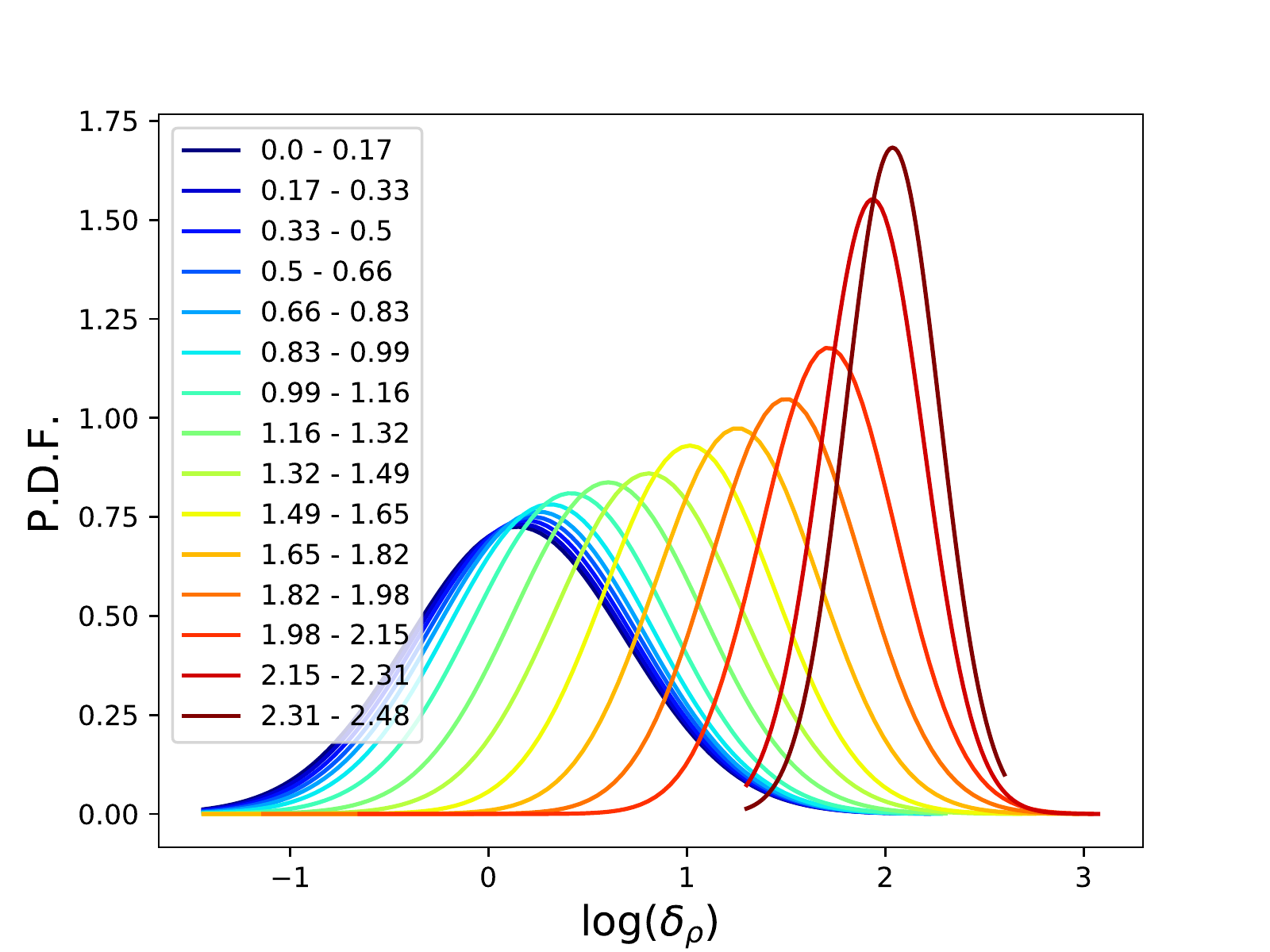}
        \caption{Probability distribution functions of the logarithm of the 
        WHIM overdensity $\delta_{\rho}$ for each of the 15 bins, with smoothing a=1.2 Mpc. In the plot legend,  we indicated the range of bins 1--15 with their
        value of $\delta_{LD}$. Notice how the distributions of the two highest--LD bins have been truncated towards the tail, to prevent a small
        number of outlying datapoints to skew the lognormal fits.}
        \label{fig:hist2}
\end{figure*}

 We then subdivide the LD range that $\delta_{LD}=1 - 300$ into 15 equally-spaced logarithmic bins
as in \cite{nevalainen2015},
and find the conditional probability distribution
of the WHIM density in each of the 15 LD bins. The choice of 
binning according to LD  values is suggested by the need to reduce the
correlation among neighboring ($\log \delta_{LD}$, $\log \delta_{\rho}$) datapoints, while retaining a sufficiently large number of bins for the sake
of an accurate determination of the two--parameter relation. In fact, both quantities are implicitly dependent on the 
underlying dark matter distribution, which drives
the formation of both the WHIM filaments and the galaxies.
To characterize the distribution of the WHIM density in each LD bin we 
 fit the distribution of WHIM overdensities $\delta_{\rho}$ with a log--normal
 model, as in \cite{nevalainen2015}, which is a more satisfactory choice than a simple normal
 model, as illustrated in Figure~\ref{fig:hist1} for few representative LD bins.
  The result of all 15 LD bins are shown in Figure~\ref{fig:hist2} and Table~\ref{table:lognorm_values}.

\begin{table}
\begin{center}
\caption{Log-normal values for each of the 15 {\bf distributions}, using smoothing $a=1.2$ Mpc.} 
\label{table:lognorm_values}
\begin{tabular}{|c|c|c|c|c|}
\hline
bin & Mean & standard deviation & number of cells \\
\hline\hline
1 & 0.14 & 0.55 & 83820 \\
\hline
2 & 0.16 & 0.54 & 109895 \\ 
\hline
3 & 0.19 & 0.54 & 148395 \\
\hline
 4 & 0.22 & 0.53 & 205519\\
\hline
5 & 0.25 & 0.52 & 291723\\
\hline
6 & 0.31 & 0.51 & 412380\\
\hline
7 & 0.42 & 0.49 & 555065\\
\hline
8 & 0.60 & 0.48 & 662312 \\
\hline
9 & 0.81 & 0.46 & 578294\\
\hline
10 & 1.01 & 0.43 & 404037 \\
\hline
11 & 1.25 & 0.41 & 228770\\
\hline
12 & 1.50 & 0.38 & 112460 \\
\hline
13 & 1.71 & 0.34 & 51394 \\
\hline
14 & 1.94 & 0.26 & 21632 \\
\hline
15 & 2.04 & 0.24 & 6899 \\
\hline
\end{tabular}
\end{center}
\end{table}

\begin{figure}
\centering
        \includegraphics[width=3.5in,angle=0]{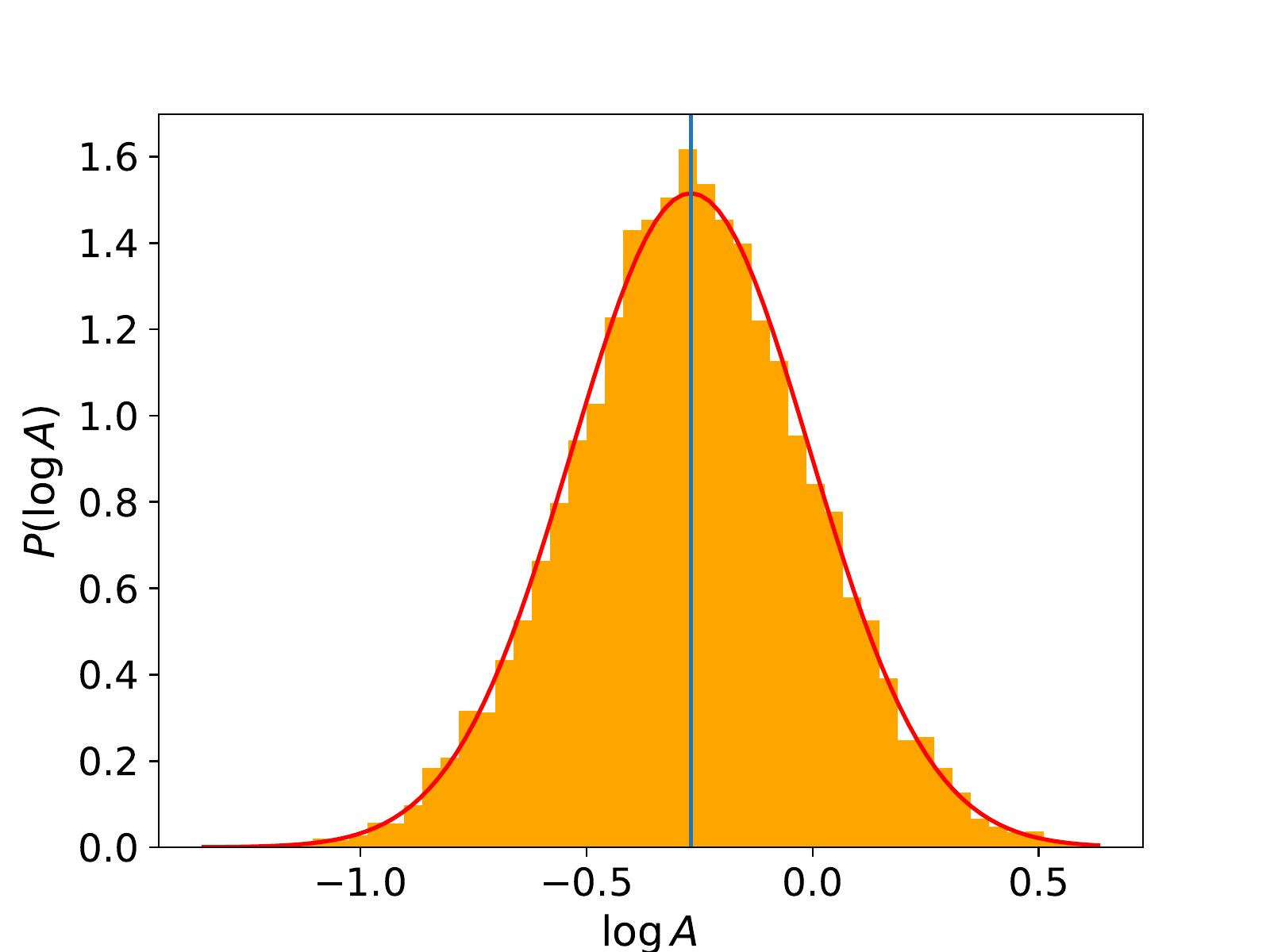}
        \includegraphics[width=3.5in,angle=0]{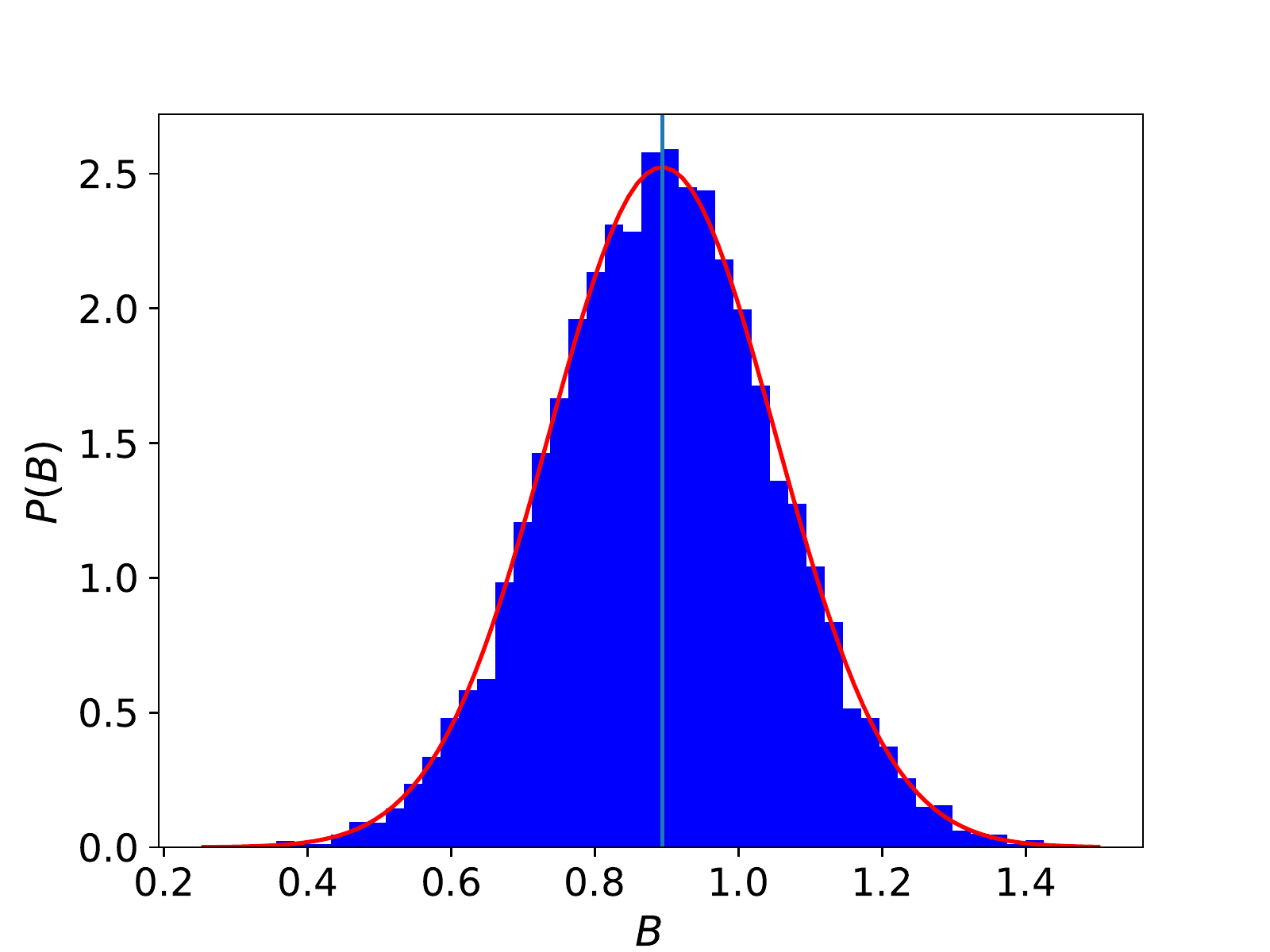}
        \caption{Probability distributions for the two parameters $\log A$ and $B$ of the WHIM--LD scaling relation (Eq. \ref{eq:logSR}) , using smoothing $a=1.2$. }
        \label{fig:histogram}
\end{figure}

\begin{figure*}
    \centering
    \includegraphics[width=6in]{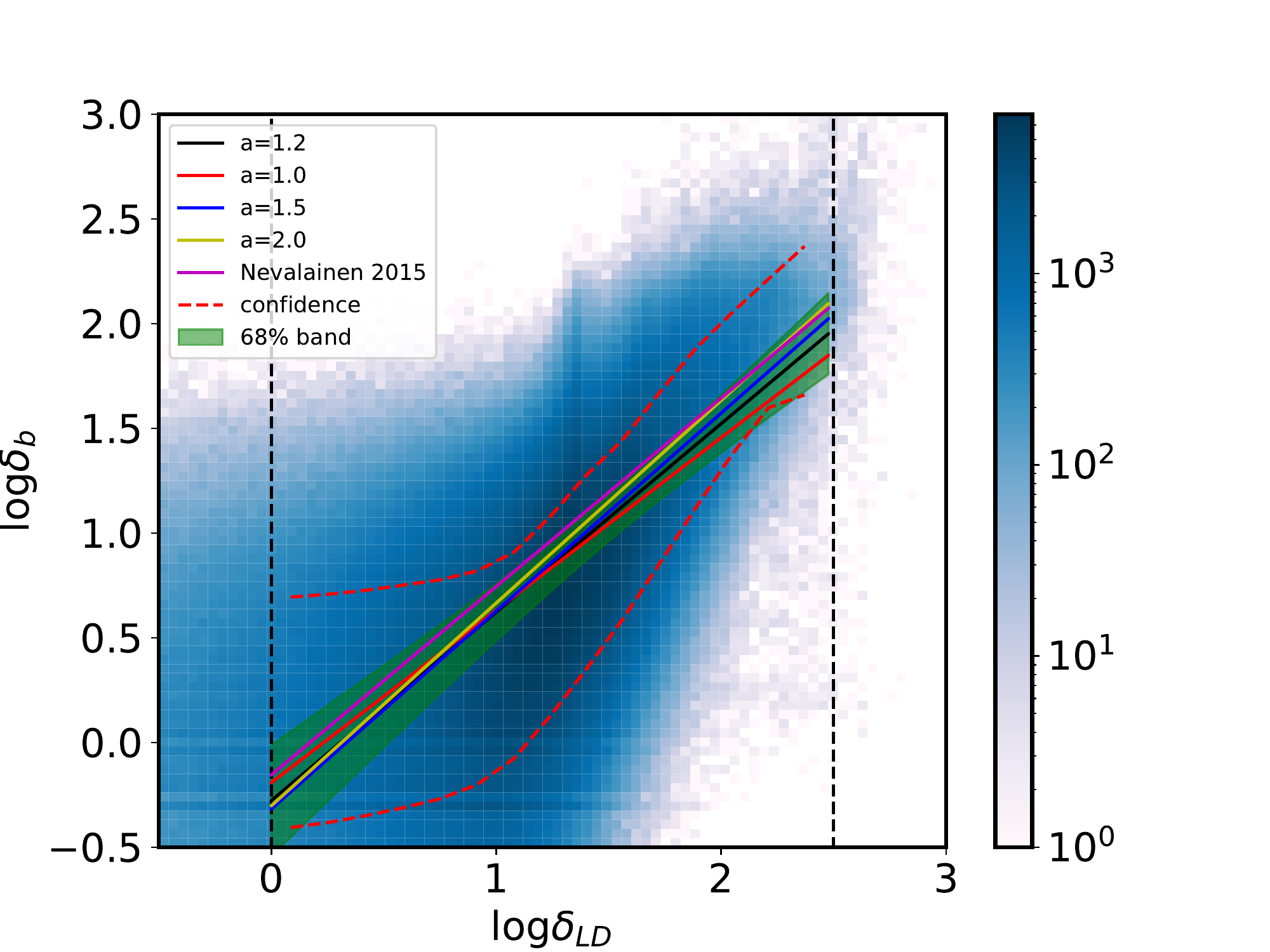}
    \caption{Best--fit scaling relation between the galaxy luminosity density (LD) and the WHIM density for the nominal smoothing of $a$=1.2 Mpc (black line), and for other values of the smoothing kernel, 1.0, 1.5 and 2.0 Mpc, indicated as solid black, red and orange lines. The scaling relation
    obtained by \protect\cite{nevalainen2015}, obtained from a different set of simulations, is also plotted for comparison in purple. The green band indicates the 68\% scatter interval of the power-law approximation to the LD - WHIM density relation. The red dashed lines indicate the 68\% scatter interval of the original data.
        }
    \label{fig:scalingRelations}
\end{figure*}

 We want to characterize the relationship between  $\log \delta_{LD}$ and
$\log \delta_{\rho}$ with the simple linear model of Eq.~\ref{eq:logSR}.
We also determine  its uncertainty or confidence band, noting that
such characterization of the best--fit uncertainties will not be used for uncertainties in the derived quantities, such as
the estimated WHIM mass, which will be done in Sect.~4.
 With the lognormal conditional probability distributions of $\log \delta_{\rho}$ for a fixed value of $\delta_{LD}$, or $f(\log \delta_{\rho}/\delta_{LD,i})$, we are ready to obtain 
the best--fit parameters for the scaling relation \eqref{eq:logSR}. 
Following \cite{nevalainen2015}, we perform a Monte-Carlo sampling of
 the  log-normal distribution of each LD bin to obtain
 a dataset of 15 datapoints, and then for each sampled dataset we
perform a simple linear regression to the 15 data points and repeat this process 10,000 times. This Monte Carlo sampling of the conditional
distributions
provides the distributions of the two parameters of the linear model, and their covariance. The distribution of the best--fit parameters $\log A$ and $B$ are normal to a very
good approximation, as shown in Figure~\ref{fig:histogram}. 
We take the mean of each parameter's distribution as the best-fit parameter value 
for our linear model, and the central 68\% confidence interval as the 1--$\sigma$ uncertainty.
This method of regression for the 15 LD bin datapoints is preferrable to a 
standard
regression of the weighted data points, since the conditional
distributions functions are not independent measurements but  correlated conditional distributions to which the standard method  of linear regression is
not directly applicable.
The choice of 15 LD bins for use in the least--squares regression is the result of a heuristic choice that lets us model
accurately the  ($\delta_{LD}$, $\delta_{\rho}$) phase space  with a simple
two--parameter linear model  that features an uncertainty (the green band in
Fig.~\ref{fig:scalingRelations}) that is  well contained within
the original scatter in the simulations (see  the red dashed curves in Fig~\ref{fig:scalingRelations}, which correspond to the data of Table~\ref{table:lognorm_values}). 
The sampling from the 15 distributions also affords a simple method to provide an estimate for the uncertainties in derivative quantities, such as the predicted
WHIM mass that is presented in Sect.~\ref{sec:observations}.

The scaling relation is illustrated in Fig.~\ref{fig:scalingRelations}, where the
confidence band is obtained as the range that contains 68\% of the
models from the 10,000 Monte Carlo
samples used for the best--fit. An equivalent
confidence band can also be obtained from an error propagation of the $\log A$ and $B$ parameters, at fixed values of the x--axis,
which takes into account the covariance between the parameters \citep{bonamente2017book}.
In the range $\log \delta_{LD}=0-2.5$, corresponding approximately to
overdensities LD$=1-300$, the scaling relation predicts a  WHIM density 
that varies by two orders of magnitude. The typical relative uncertainty in the prediction
of the WHIM density at a fixed LD is given by the
size of the green band, which is less than $\nicefrac{1}{2}$ dex at all values of the luminosity density. This level of uncertainty  will be 
reflected in the uncertainty in predicted WHIM masses when using the scaling
relation. Notice how the best-fit scaling relation for other values of
the smoothing parameter in the range $a=1-2$~Mpc fall comfortably within the 
1--$\sigma$ uncertainty of the $a=1.2$~Mpc scaling relation. 

\begin{table}
\begin{center}
\caption{Best fit parameter values for selected smoothing  scales.}
\label{table:ParamFits}
\begin{tabular}{|c|c|c|c|c|}
\hline
smoothing [Mpc] & $\log$ A & $\sigma_{\textrm{log A}}$ & B & $\sigma_{\textrm{B}}$\\
\hline\hline
0.5 & 0.42 & 0.25 & 0.43 & 0.16 \\
\hline
1.0 & -0.19 & 0.26 & 0.82 & 0.16 \\ 
\hline
1.2 & -0.27 & 0.27 & 0.89  & 0.16 \\
\hline
1.5 & -0.32 & 0. 27 & 0.94 & 0.16 \\
\hline
2.0 & -0.30 & 0.28 & 0.97 & 0.17 \\
\hline
\end{tabular}
\end{center}
\end{table}

\subsection{Systematic biases and errors}

\label{sec:systematics}
\begin{figure}
    \centering
    \includegraphics[width=3in]{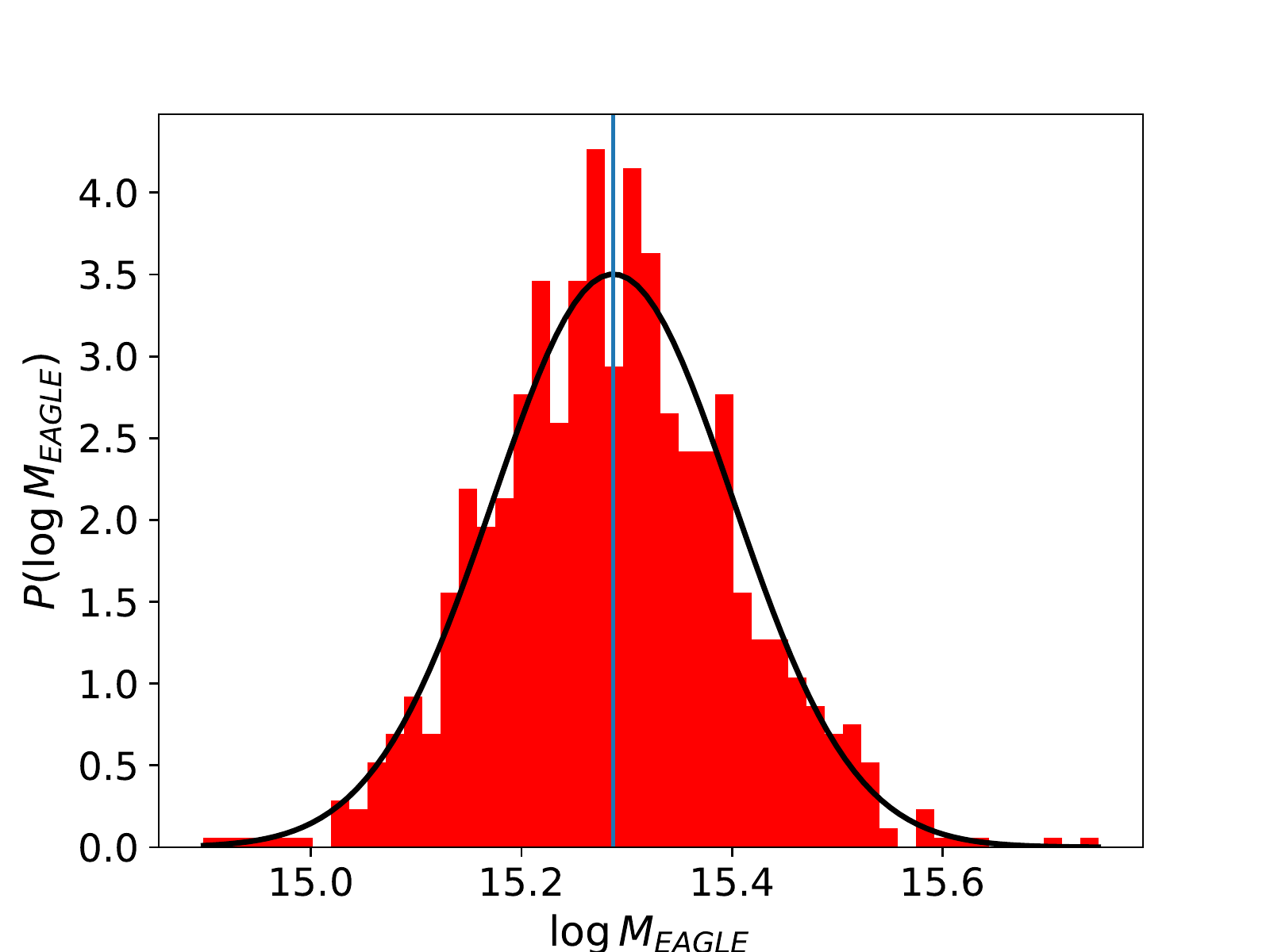}
    \includegraphics[width=3in]{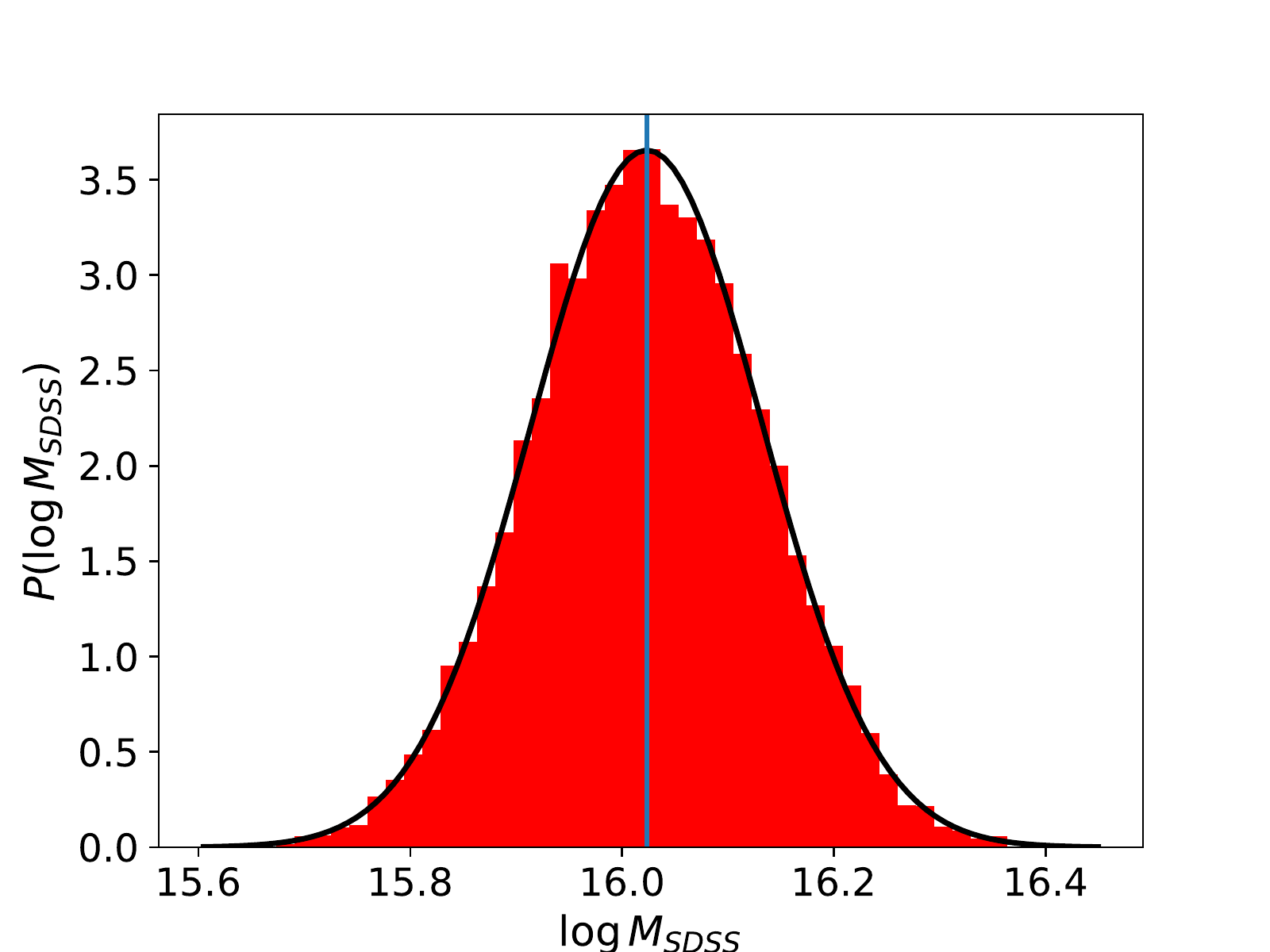}
    \caption{(a) Distribution of the predicted EAGLE WHIM mass in logarithmic scale, in units of $M_{\odot}$. The mean is 15.28, which converts to $1.9 \times 10^{15} M_\odot$. The mass distribution was obtained using all parameter values obtained from the Monte-Carlo simulation. Overplotted is the best--fit normal model. 
    (b) Distribution of the predicted SDSS WHIM mass, with a mean of $\log M=16.02$.  The mass distribution was obtained the same way as the EAGLE mass, using all parameter values obtained from the Monte-Carlo simulation. Overplotted is the best--fit normal model.}
    \label{fig:EAGLE_mass_dist}
    \label{fig:SDSS_mass_dist}
\end{figure}

The true WHIM filament mass in the entire EAGLE volume is $2.15 \times 10^{15}$~\msun\ (see Fig.~\ref{fig:EAGLE_mass_dist}), as obtained from a direct sum of all baryon particle masses
in Bisous filaments and outside of $r_{200}$, and in the 
temperature range $\log T(K) = 5-7$, regardless of galaxy luminosity. To calibrate the 
LD-WHIM density scaling relation
to reconstruct WHIM masses based on the galaxy LD, it is necessary to take into
account the key choices made in the analysis and estimate biases and systematic errors
associated with them. In the following we investigate the effects of two key elements of the
analysis, and how we correct for their biases and systematic errors.

\subsubsection{Choice of smoothing kernel and LD range}
\begin{figure}
    \centering
    \includegraphics[width=3in]{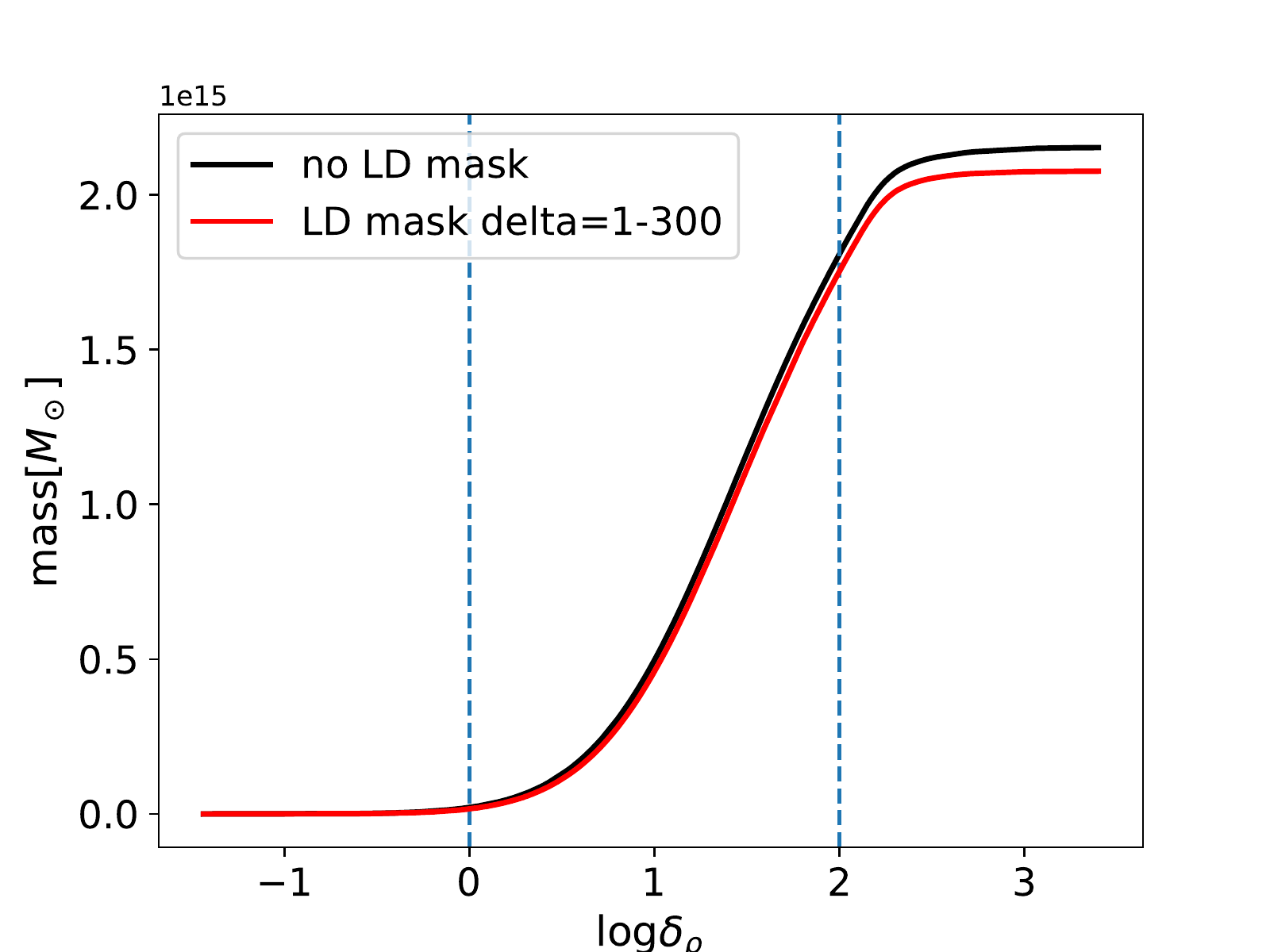}
    \caption{Cumulative distribution of the true EAGLE WHIM mass in filaments, for a smoothing parameter $a=1.2$~Mpc.}
    \label{fig:CumulativeMass}
\end{figure}

The choice of a LD filter and the parameter $a$ of the smoothing kernel are two
key parameters used in the analysis. Table~\ref{table:WHIMmass} shows 
the EAGLE WHIM masses using the LD range $\delta_{LD}=1-300$ (labeled "LD mask"), which is a function of the smoothing parameter $a$, and the predicted WHIM mass
using the same cells to which the scaling relation was applied ("Prediction from SR"). The uncertainty in the predicted masses from the scaling relation
were obtained in the same manner as the green band from Fig.~\ref{fig:scalingRelations}, i.e., by using 10,000 Monte Carlo 15--point samples.
The results of the reconstruction of WHIM masses from the LD-WHIM scaling relations
of Table~\ref{table:WHIMmass} show that the nominal scaling relation for $a=1.2$
does an excellent job at reconstructing the WHIM mass, with a small
bias of $\leq -10$\% ($1.89 \times 10^{15}$~M$_{\odot}$ versus $2.08\times 10^{15}$~M$_{\odot}$), 
when both the true WHIM  mass and the one inferred by the scaling relations are limited to the 
LD range of $\delta_{LD}=1-300$ (see also Fig.~\ref{fig:CumulativeMass} ).
Although the agreement between the true EAGLE WHIM mass and that reconstructed from the scaling relation is  well within the statistical uncertainties of the 
scaling relation, the bias is \emph{systematic}, in that the scaling relation always produces a \emph{lower} estimated mass than the true one. As a result, such $\sim 10$\% systematic bias
must be \emph{corrected} by boosting the inferred WHIM mass by $+10$\%.~\footnote{
For larger values of the smoothing parameter, e.g. $a=2$~Mpc, the reconstruction
has a larger bias of up to -25\%, which is attributable to a larger fraction of the
LD being moved outside of the WHIM filament regions by the wider smoothing kernel. Our analysis, however,
does not support the use of such large smoothing kernels, and this larger bias is only reported
as a caution against using too wide smoothing kernels.}
Likewise, the effect of limiting the LD range to $\delta_{LD}=1-300$ provides another systematic
bias of approximately $-5$\% ($2.15 \times 10^{15}$~M$_{\odot}$ versus $2.08\times 10^{15}$~M$_{\odot}$), also
qualitatively consistent with Figure~\ref{fig:CumulativeLD}. This is also a systematic bias
that needs to be corrected with an additional $+5$\% boost of the estimated mass. The combined effect of
these two systematic biases is that the estimated masses must be corrected by $+15$\%, as also
illustrated in Table~\ref{tab:systematics}.

\begin{table*}
\begin{center}
\caption{Estimated WHIM mass in the EAGLE simulation using the various scaling relations.}
\label{table:WHIMmass}
\begin{tabular}{|c|ccc |c|}
\hline
$a$ (Mpc) & \multicolumn{2}{c}{EAGLE WHIM Masses ($\textrm{M}_\odot$)} & Prediction from SR\\
          & No LD mask & LD mask & \\
\hline\hline
0.5 & $2.15 \times 10^{15}$ & $1.31 \times 10^{15}$ & $1.03 \pm 0.19 \times 10^{15}$ \\
1.0 & $2.15 \times 10^{15}$ & $2.02 \times 10^{15}$  & $1.85 \pm 0.39 \times 10^{15}$  \\ 
1.2 & $2.15 \times 10^{15}$ & $2.08 \times 10^{15}$ & $1.89 \pm 0.41 \times 10^{15}$\\
1.5 & $2.15 \times 10^{15}$ & $2.12 \times 10^{15}$ & $1.8 \pm 0.41 \times 10^{15}$ \\
2.0 & $2.15 \times 10^{15}$  & $2.14 \times 10^{15}$  & $1.6 \pm 0.39 \times 10^{15}$ \\
\hline
\end{tabular}
\end{center}
\end{table*}

\subsubsection{Choice of Bisous method of filament detection}
An additional source of systematic bias and error is the method of detection of WHIM filaments.
As shown in \cite{tuominen2021}, the Bisous method successfully finds the
vast majority ($\sim 87$~\%) of the missing WHIM baryons outside
the virial radii of galaxies and in the $\log T(K) \geq 5.5$ range. 
This large 'capture' rate, using the \cite{tuominen2021} terminology, 
indicates that the Bisous method leaves out a fraction of 
WHIM gas
in the outskirts of the thickest filaments
that are less correlated to the galaxy luminosity, and therefore not included in the estimate
of the WHIM mass from the scaling relation.  Although there is visual evidence that
several filaments are somewhat \emph{wider} than the Bisous--identified volumes, especially in the densest regions where the Bisous method does not have the flexibility to adapt its
size (see discussion in \citealt{tuominen2021}), we do not assess a correction to our estimates for this effect.

Moreover, \cite{tuominen2021} compared the total hot WHIM mass in the EAGLE simulation within filaments detected with the Bisous and the NEXUX+ method, which is a different method of filament identification,
and estimated a $\pm 10$\% difference between the two methods. 
It is therefore reasonable to assess 
a systematic error of  $\pm 10$\%, in order to correct for the
effects of the use of the Bisous method.
A summary of the major sources of systematic error in the estimate of
the WHIM masses from the LD-WHIM scaling relation is provided in Table~\ref{tab:systematics}.

\subsubsection{Other sources of systematic errors}

 There are several other possible sources of systematic errors in the estimation
 of WHIM masses from the LD-WHIM density scaling relation. 
 The comparison between the scaling relation derived from the earlier simulations used in \cite{nevalainen2015} and the
 present EAGLE simulation is provided in Fig.~\ref{fig:scalingRelations}. Despite the
 numerous differences between the two numerical simulations, which range from
 feedback methods to resolution etc., the two scaling relations are in statistical agreement.
 
 When the EAGLE WHIM mass is reconstructed using the 
 \cite{nevalainen2015} scaling relation, instead of the
  EAGLE--derived relation, the mass is
   $2.5 \pm 0.5\times10^{15} \textrm{M}_\odot$, which is approximately 30~\% larger that the
   one estimated in Tab.~\ref{table:WHIMmass} according to the present scaling relation, although in statistical agreement within the uncertainties of the EAGLE simulations.
 This difference in reconstructed masses
 provides an estimate of the effect of other sources
 of systematic errors associated with the derivation of the WHIM--LD scaling relation. We therefore add a $\pm 30$~\% systematic error
 to mass estimates from the LD-WHIM density scaling relation method as a means to address uncertainties associated with other prescriptions of
 the numerical simulations.

\begin{table}
    \centering
    \begin{tabular}{l|c}
    \hline
    \hline
    Source of systematic bias and error & Effect on WHIM mass\\
    \hline
    \multicolumn{2}{c}{Systematic biases}\\
    Choice of LD boundaries & $-5$\% \\
    Use of $a=1.2$~Mpc smoothing kernel &  $-10$\%  \\
    \multicolumn{1}{c}{\hrulefill} & \\
    Summary of correction for systematic biases & $+15$\% \\
    \hline 
    \multicolumn{2}{c}{Systematic  uncertainties}\\
    Choice of filament detection method & $\pm 10$\% \\
    Other choices in the numerical simulations & $\pm 30$\% \\
    \hline
    \hline
    \end{tabular}
    \caption{Major sources of systematic bias and errors associated with the LD-WHIM
    scaling relation. All biases systematically underpredict the WHIM mass, and therefore 
    a positive correction is required to offset this bias.}
    \label{tab:systematics}
\end{table}



\section{Application of the scaling relation to SDSS galaxy filaments}
\label{sec:observations}
\begin{figure}
    \centering
    \includegraphics[width=3in]{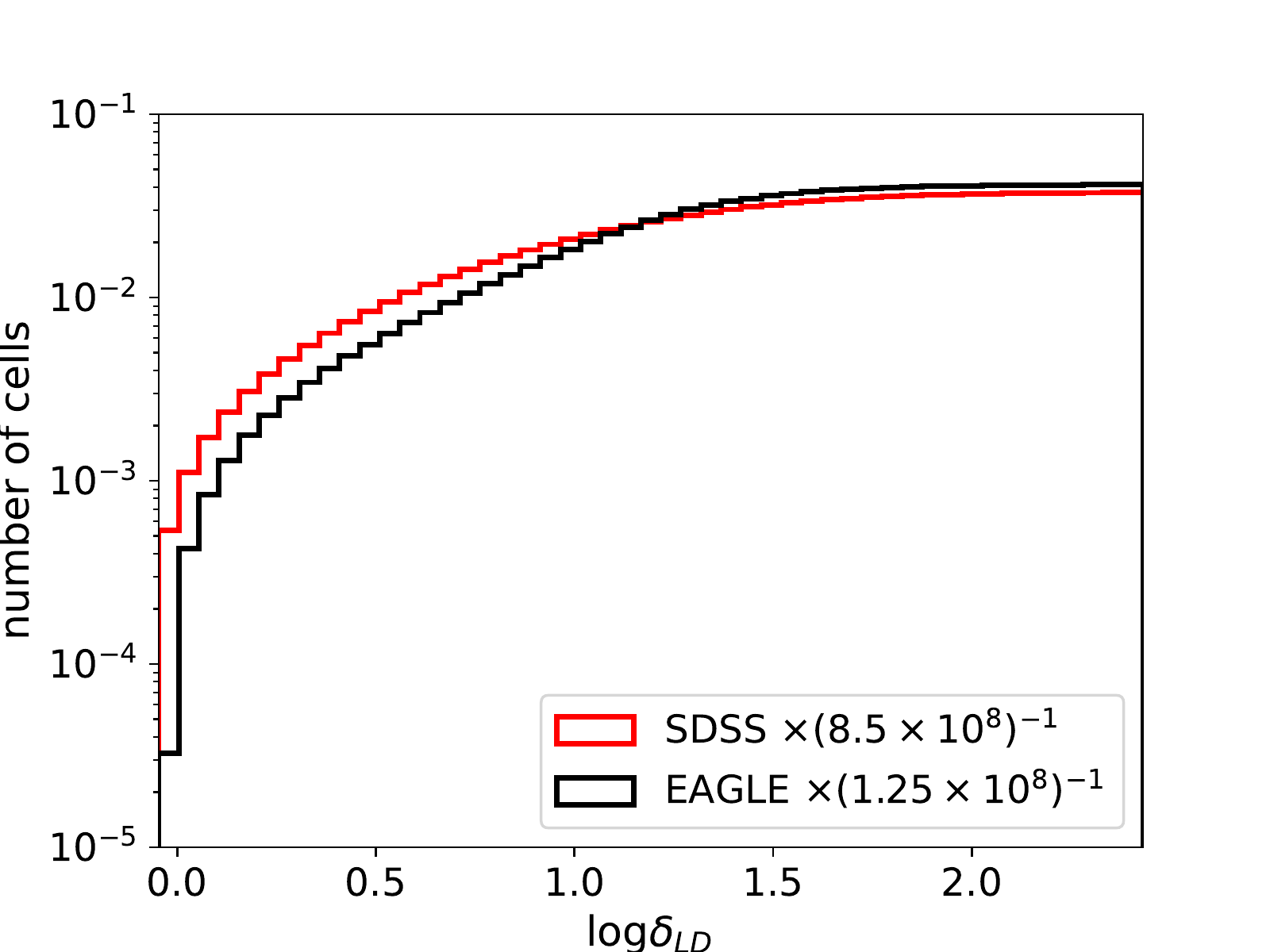}
       \caption{Cumulative distribution of the density of LD cell counts in filaments, for EAGLE and SDSS. The number of cells
       was rescaled by the volume of each data set, indicating that EAGLE contains approximately $15$\% more LD cells, per unit volume, than SDSS.}
    \label{fig:Cumulative_LD_cells}
\end{figure}

In order to examine the 
 performance of our simulation-based LD-WHIM density relation with  real--life data, we applied it to the  SDSS DR12 galaxy data \citep{york2000,ahn2014}
that was previously analyzed by \cite{tempel2014a, tempel2016} and \cite{kuutma2020} to identify
filament spines with the \textit{Bisous} method (see Table~\ref{tab:SDSS} for the key features of the SDSS data used in this paper).
The data correspond to the Legacy Survey, for a 7,221
contiguous square degrees or 17.5\% of the full
sky \citep{martinez2009}. 
For the present study, we 
consider only galaxies at redshifts $z=0.02-0.05$, corresponding
to a distance of approximately $88-220$ Mpc for $H_0=67.8$~\kmsMpc,
as a volume--limited representative sample of the local universe. A description of the selection of this
volume--limited sample can be found in \cite{tempel2014b}.
In brief, the choice of a low--redshift limit avoids the use
of a local void that is underpopulated, and the
upper limit mitigates the loss of low--luminosity galaxies,
and the associated filaments, at
higher redshift. This sample of galaxies is therefore
intended to be representative of the low--redshift universe, and its
volume--limited nature makes the ensuing cosmological inferences
straightforward. While the EAGLE and SDSS volumes differ by a factor of ~8, the filament volume filling fractions are very similar,  $\sim$5\%.
Moreover, Fig.~\ref{fig:Cumulative_LD_cells} shows that EAGLE and SDSS
contain a comparable number of LD cells, after accounting for the volume
difference.

\begin{figure*}
    \centering
    \includegraphics[width=6in]{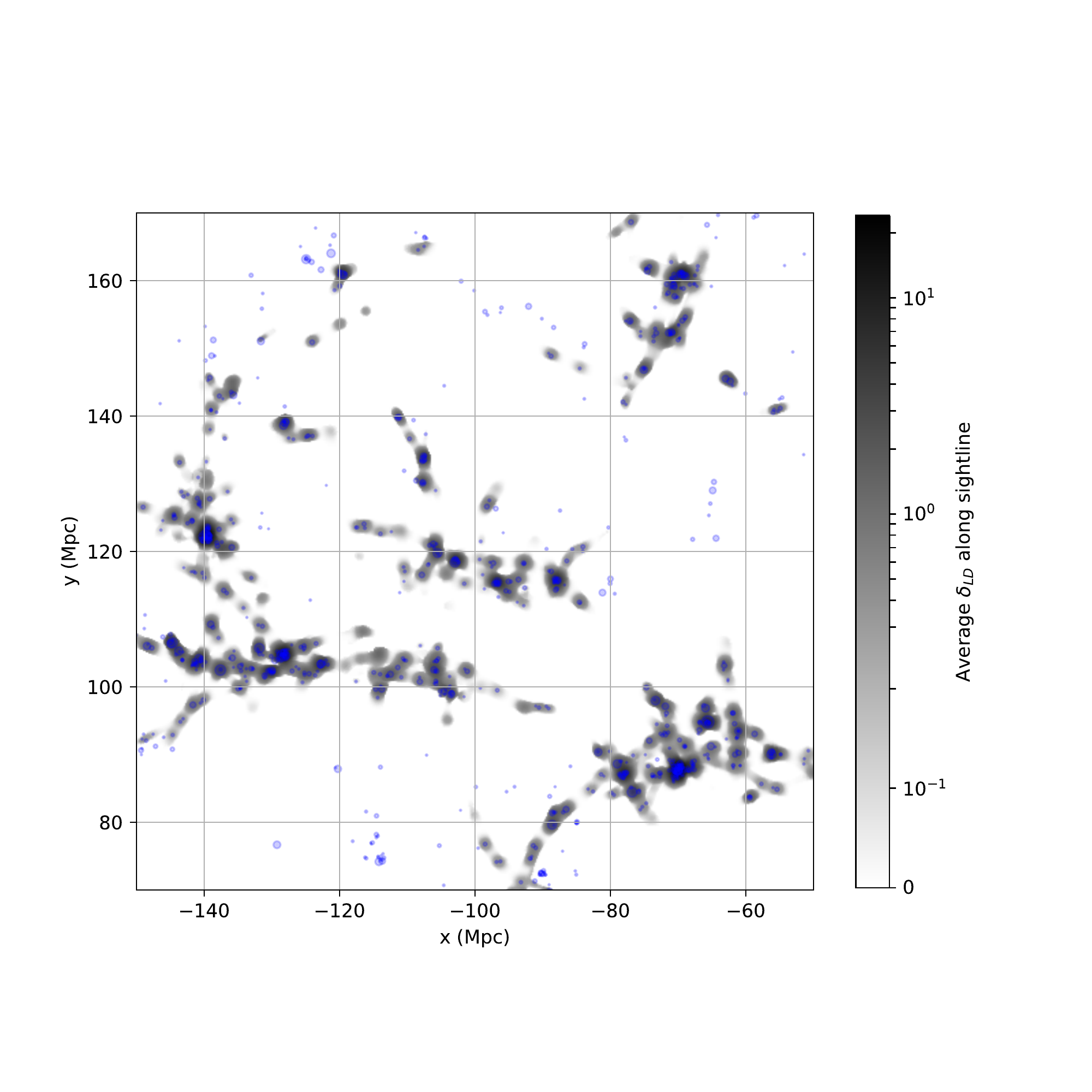}
    \caption{Image of a $100\times 100$ Mpc$^2$ portion of the SDSS, with LD intensities averaged along a depth of 5~Mpc. The grey--scale cells represent the smoothed LD in the \textit{Bisous} filaments, and the
    blue dots represent the SDSS galaxies (larger circles represent higher $r$--band luminosity) used to identify the filaments. }
    \label{fig:sdssImage}
\end{figure*}

We smoothed the measured $r$--band galaxy luminosity with our adopted  $a=1.2$~Mpc kernel (see Fig.~\ref{fig:sdssImage} for a slice of the LD field).
For a given pair of $A$ and $B$ values of the scaling relation (Eq. \eqref{eq:Mfil}) of the Monte Carlo samples (see Section X), we then turned each LD value into WHIM density, and computed the total predicted WHIM mass in the SDSS filaments using
\begin{equation}
\log M_{\mathrm{fil, SR}} = \sum_{i=1}^{N_{\mathrm{SDSS}}}
 (\log A + B \log \delta_{LD,i}) + \log \rhob + \log \Delta V_{\mathrm{SDSS}} 
 \label{eq:Mfil}
 \end{equation}
where $\delta_{LD,i}$ is the measured luminosity overdensity in each of
the $N_{\mathrm{SDSS}}$
cell of fixed volume $\Delta V_{\mathrm{SDSS}}=0.2^3$~Mpc$^{3}$ within Bisous filaments.
We repeated this for the  10,000 Monte Carlo samples obtaining a distribution of the predicted total WHIM mass in Bisous filaments in the SDSS volume (see Fig.~\ref{fig:SDSS_mass_dist}).
This yields WHIM mass of $\log M_{\mathrm{fil, SR}} = 16.02\pm 0.11$, where the best--fit value and there error
are, respectively, the mean and standard deviation of the distribution.
This value corresponds to approximately 
\[M_{\mathrm{fil, SR}}=1.054\pm 0.265\times 10^{16}~\mathrm{M}_{\odot}.\] 
This mass represents the WHIM in filaments as estimated from the LD-WHIM density
scaling relation. However, prior to its use for cosmological inferences, it is
necessary to remove the biases discussed in Sect.~\ref{sec:systematics}.
Since the two sources of systematic bias, i.e., restriction to the
$\delta_{LD}=1-300$ range and the LD smoothing which removes a fraction of
the luminosity from the filament volume (see Table~\ref{tab:systematics}), require
a combined correction by {\bf $+15$\%}, the debiased WHIM mass inferred by our analysis 
becomes
\[M_{\mathrm{fil, SR,corr}}=1.21\pm 0.30\times 10^{16}~\mathrm{M}_{\odot}\]

The contribution of our predicted baryon content in the SDSS volume V$_{SDSS}$ to the cosmic baryon budget is obtained as
\begin{equation}
\Omega_{\mathrm{b,LD}}= \dfrac{1}{\rhob} \dfrac{M_{\mathrm{fil, SR,corr}}}{V_{\mathrm{SDSS}}}. 
       \label{eq:omegab}
\end{equation}
The fraction of  sky area covered by the SDSS's main footprint is 7221/41252.96 = 0.175,
and therefore the volume between the two co--moving radial distances $d_1=88$~Mpc and $d_2=220$~Mpc
is given by
    \[ V_{\mathrm{SDSS}} = 0.175 \times \dfrac{4}{3} \pi \left(d_2^3 - d_1^3\right) =  6.84  \times 10^{6}~\textrm{Mpc}^3. 
    \]
In terms of $\Omega_b$, our prediction for the SDSS WHIM density translates to 
\begin{equation}
\Omega_{\mathrm{b,LD}}=  \left(0.31 \pm 0.07 {\mathbf\pm{0.12}}\right) \, \Omega_b,
\label{eq:OmegabSDSS}
\end{equation}
where the best--fit value is bias--corrected, and 
the additional {\bf sources} of systematic error described in Sect.~\ref{sec:systematics} {\bf were} added to the statistical errors associated with the determination of the scaling relation.

This result indicates that if our EAGLE-based LD-WHIM density relation accurately describes the real universe, the baryon content of the SDSS volume traceable by our methods corresponds to a substantial fraction of the expected cosmic baryon density. Additionally, our result is consistent with the current estimates of the observationally missing baryons \citep{shull2014, danforth2016}. This implies that our updated relation, together with the Bisous filament finder and the LD field method, may be a useful tool for the missing baryon searches.

\begin{table*}
    \centering
        \caption{Key parameters of the SDSS DR12 Legacy Survey galaxies used in this paper and \protect\cite{tempel2014a}.}
    \begin{tabular}{l|c}
    \hline
         Redshift boundaries& 0.02--0.05 ({88--220} Mpc)  \\
         Contiguous sky area covered & 7,221 sq. deg. \\
         Number of galaxies with $M_r \leq -19.5$ & 55973 \\ 
         Average <LD> & $1.1\times10^8$~$L_{\odot}$~Mpc$^{-3}$ \\
         Fraction of volume occupied by WHIM filaments in SDSS and EAGLE & 5.0 \% \\
         Fraction of volume occupied by WHIM filaments in SDSS ($ \delta_{LD}=1-300$) & 3.7 \% \\
         Fraction of volume occupied by WHIM filaments in EAGLE ($ \delta_{LD}=1-300$)   & 4.2\% \\ 
         \hline
    \end{tabular}

    \label{tab:SDSS}
\end{table*}

\section{Discussion and conclusions}
\label{sec:conclusions}

This paper has presented an update to the
\cite{nevalainen2015} 
 method to trace cosmic baryons with
the use of optical light from galaxy surveys, with the use of EAGLE
numerical simulations.
The technique studies the correlation between
the optical luminosity density of galaxies and the WHIM density
in galaxy filaments that are identified by the \textit{Bisous} method,
with the goal to predict WHIM masses in galaxy filaments.
The first result of our analysis is that the EAGLE simulations
predict a strong correlation between the WHIM gas density and the 
smoothed galaxy luminosity density in filaments, confirming the earlier
funding of \cite{nevalainen2015} based on lower resolution simulations. This is consistent with the scenario whereby both the diffuse WHIM gas and the stellar component in the galaxies trace the same underlying dark matter density field, enabling the usage of the relatively easily observable galaxy luminosities in spectroscopic surveys as a tracer of the more hard--to--detect WHIM.

We then established a new LD-WHIM density scaling relation based on the EAGLE simulations. 
The 68\%~scatter in the scaling relation
at a fixed value of the luminosity density is less than $\nicefrac{1}{2}$ dex,
throughout the range of the luminosity densities under consideration (i.e,
overdensities of $\delta_{LD}=1-300$). This is a manageable
amount of scatter, and it endows the scaling relation with the power to make
meaningful predictions for the WHIM mass, based on the unrelated luminosity
density observable. 


We applied the scaling relation
to a sample of low--redshift galaxies ($0.02 \leq z \leq 0.05$) 
from the SDSS DR12 Legacy data, for which
\cite{tempel2014a} had already identified Bisous filaments, in an area that covers $\sim 17.5\%$ of the sky. 
Our analysis predicts that the optical luminosity of filaments, outside of the virial radii
of the galaxies, traces approximately \OLD$\simeq 31 \pm10$\% of 
the cosmological density of baryons \omegab. 
This is consistent with the current estimates of the observationally missing baryons.
Thus, assuming that our EAGLE-based relation accurately captures the real universe,  our method has the potential to address the missing baryon search.

The main motivation for this LD-WHIM scaling relation is to make predictions
of the WHIM density for filaments detected in large portions of the sky by, e.g., the SDSS
and by upcoming 4MOST surveys \citep{winkler2020}, in order to identify the most promising sight--lines
for follow--up observations in the FUV and  X--rays (e.g. with Athena, \citealt{barret2020}). In this case,
the LD-WHIM scaling relation may become to tool of choice to 
implement a synergy between the relatively inexpensive optical observations of galaxies
and the more challenging FUV and X--ray follow-ups, which can be used to confirm the nature of the WHIM only for the sight--lines with the largest predicted column density of WHIM.

\section*{Data Availability}
The EAGLE data underlying this article are available from the Virgo Consortium at  http://virgodb.dur.ac.uk/ \citep{mcalpine2016,eagle2017}, and from the SDSS project at https://www.sdss.org/ \citep[for DR12, see][]{SDSSDR12}. Processed data
and tables presented in the paper are accessible upon request to the authors.

\bibliographystyle{mnras}
\bibliography{max}

\bsp	
\label{lastpage}
\end{document}